\documentclass[aps,reprint,prb,floatfix,nofootinbib]{revtex4-2}
\usepackage{hyperref}
\hypersetup{
  breaklinks=true, 
  colorlinks=true, 
  pdfusetitle=true, 
}
\usepackage{float}
\usepackage{epsfig}
\usepackage{graphicx}
\usepackage{amsmath}
\usepackage{amsfonts}
\usepackage{amssymb}
\usepackage{bm}
\usepackage{bbold}
\usepackage{epsfig}
\usepackage{graphicx}
\usepackage{color}
\usepackage{pictex}
\usepackage{mathtools}
\usepackage{extarrows}
\usepackage{lipsum}
\usepackage{footmisc}
\newcommand{\s}{\scriptscriptstyle}
\begin{document}
\title{Disentanglement, disorder lines, and  Majorana edge states in a solvable quantum chain}
\author{Gennady Y. Chitov}
\affiliation{Institut quantique \& D\'{e}partement de physique, Universit\'{e} de Sherbrooke,
Sherbrooke, Qu\'{e}bec, J1K 2R1 Canada}
\author{Karun Gadge}
\affiliation{School of Basic Sciences,
Indian Institute of Technology Mandi, Disst. Mandi 175075, India}
\affiliation{Institute for Theoretical Physics, Georg-August-University G\"{o}ttingen, Friedrich-Hund-Platz 1, D-37077 G\"{o}ttingen, Germany}
\author{P. N. Timonin$^\dagger$}
\affiliation{Physics Research Institute,
Southern Federal University, 194 Stachki ave.,  Rostov-on-Don, 344090 Russia}
\date{\today}

%
%
\begin{abstract}
We study the exactly solvable 1D model: the dimerized $XY$ chain with uniform and staggered transverse fields,
equivalent upon fermionization to the noninteracting dimerized Kitaev-Majorana chain with modulation.  The model has
three known gapped phases with local and nonlocal (string) orders, along with the gapless incommensurate (IC) phase
in the $U(1)$ limit.  The criticality is controlled
by the properties of zeros of model's partition function, analytically continued onto the complex wave numbers.
In the ground state they become complex zeros of the spectrum of the Hamiltonian. The analysis of those roots yields the
phase diagram which contains continuous quantum phase transitions and weaker singularities known as disorder lines (DLs)
or modulation transitions. The latter, reported for the first time in this model, are shown to occur in two types:
DLs of the first kind with continuous appearance of the IC oscillations,  and DLs of the second kind corresponding to a
jump of the wave number of oscillations. The salient property of zeros of the spectrum is that the ground
state is shown to be separable (factorized) and the model is disentangled on a subset of the DLs.
From analysis of those zeros we also find the Majorana edge states and their wave functions.
\end{abstract}
\maketitle

%
%
%
\section{Introduction}\label{Intro}
%
%
%
%

Arguably the most rigorous and fundamental approach to study phase transitions was pioneered by Yang and Lee \cite{YangLee:1952,*LeeYang:1952}.
They related transitions to zeros of model's partition function, so the theory is applicable whatever the nature of order parameter or
symmetry breaking is.  The original analysis of Yang and Lee of the statistical mechanics of the Ising model was further extended for various
models, including those out of equilibrium. See, e.g. papers \cite{Fisher:1965,Fisher:1980,Matveev:2008,TongLiu:2006,Bena:2005,Wei:2014}, and more references in there.
The requirement of the Lee-Yang zero of the partition function at the critical point of continuous thermal transition becomes the condition for zero of the spectrum (gap closure) at the quantum phase transition point. In both cases the zeros must occur in the real range of physical parameters.

In 1970 Stephenson \cite{Stephenson-I:1970,*Stephenson-II:1970,Stephenson:1970PRB} found a new type of transitions in classical Ising models which he called ``disorder lines" (DLs). The transition consists in modulation of the monotonic exponential decay of the correlation functions  by incommensurate (IC) oscillations.  At zero temperature similar modulation was later found by Barouch and McCoy \cite{McCoyII:1971} in the $XY$ quantum chain with transverse field. Disorder lines, or  modulation transitions are quite general phenomena occurring in a large variety of models  \cite{Nussinov:2011,Nussinov:2012,Salinas:2012,Chitov:2017PRE,Ogilvie:2020,Chitov:2021}. Whether one can define the modulation as a transition is a subtle issue \cite{Ruelle:1974}, since DL is a weak feature lacking identifiable divergencies in the derivatives of thermodynamic potential. A non-analyticity is manifested in the behavior of the correlation length, which remains finite but demonstrates a cusp at the DL point.

Important progress was reported recently in Refs.~\cite{Chitov:2017PRE,Chitov:2021}, where the origin of DLs was understood within the Lee-Yang formalism: 
the DLs and the IC oscillations were related to the complex conjugate zeros of the
partition function. The conventional phase transitions are well understood as occurrence of zero(s) $z$ of the partition function $\mathcal{Z} (z)=0$ in the \textit{real range of physical parameters}: temperature, field, coupling, wave number, etc.  Here $z$ stands for some conveniently chosen  function of physical quantities, like e.g., fugacity, $e^{ik}$, spectrum of the transfer matrix.  The DL corresponds to the pair of merging complex conjugate roots $\mathcal{Z}(z)=\mathcal{Z}(z^*)=0$ occurring at \textit{complex values of physical parameters}. For instance, the infinite cascades of DLs found in the classical Ising chain \cite{Chitov:2017PRE}, correspond to zeros of the partition function $\mathcal{Z}(h)$ in the range of complex magnetic field $h \in \mathbb{C}$; modulations revealed in various free fermionic models in the spatial dimensions $d=1,2,3$ correspond to zeros \footnote{\label{ZPFnote} The zeros are defined for the partition function of the finite-size system, before the thermodynamic limit is taken \cite{TongLiu:2006}.} of the partition function $\mathcal{Z}(\mathbf{q})$ in the range of complex wave vectors $|\mathbf{q}| \in \mathbb{C}$ \cite{Chitov:2021}.

An interesting aspect of the critical properties on the DL is factorization of the ground state found in the transverse field $XY$ quantum chain \cite{McCoyII:1971,Franchini:2017} leading to constant correlation functions, the hallmark of disentanglement. Since the possibility to interpret the
disentanglement point as a transition of some kind  is actively studied in the literature \cite{Wolf:2006,Amico:2006,Illuminati:2013,WeiGold:2005,*WeiGold:2011}, it is a very pertinent problem to clarify the relation between the zero-temperature limit of the Lee-Yang zeros 
\begin{equation}
\label{LYZT0}
  \mathcal{Z} (z)=0~ \mathrm{at} ~T \to 0
\end{equation}
and disentanglement. It was shown in the recent analysis \cite{Chitov:2021} of the well-known quantum $XY$ chain in the uniform transverse field, that the DL which is also the line of disentanglement in that model, correspond to the line of the merging complex roots of \eqref{LYZT0}. The important question is to clarify whether the merging complex roots (DLs) necessarily signal the factorized ground states (disentanglement) in other models.

To this end we analyse in this paper the dimerized quantum $XY$ spin chain in the uniform and staggered transverse fields. This is an exactly-solvable free fermionic model with rich phase diagram \cite{Perk:1975,DuttaTIM:2015,Chitov:2019}. The DLs and disentanglement in this model were not studied before. The
DL as a modulation transition is also manifested in the IC oscillations of the wave function of the Majorana edge states, as reported in \cite{Karevski:2000} for the homogeneous transverse $XY$ chain. After the work by Kitaev on that model \cite{Kitaev:2001}, the Majorana modes in quantum condensed matter systems became a very active topic of research, for reviews, see e.g., \cite{Alicea:2012,Valkov:2022}. The existence of those edge modes in quantum chain was linked to the localization of zeros of the \textit{resolvent of the transfer matrix} on the complex plane  in earlier studies \cite{DeGottardi:2011,DeGottardi:2013b}. The resolvent, as was shown earlier for the special case \cite{Chitov:2018}, does not provide an independent information, since the eigenvalues of the transfer matrix for the localized Majorana states are zeros of the spectrum, i.e., the zero-temperature limit of the roots \eqref{LYZT0}, lying within unit circle on the complex plane. This result, identifying the eigenvalues of the transfer matrix and the complex roots 
of the spectrum, is shown to be valid for the more general model studied in this work.

The main goal of the present study is to demonstrate from the systematic analysis of zeros $z_\pm$ of the spectrum  of the exactly solvable model \eqref{XYHam}, which are also the zero-temperature limit of the Lee-Yang zeros \eqref{LYZT0}, that $z_\pm$ control the critical points, modulation transitions, disentanglement, winding numbers, and the Majorana edge modes.

The rest of the paper is organized as follows:
In Sec.~\ref{Model} we introduce the model to be studied and give an account on its spectrum and the ground-state phases.
In Sec.~\ref{DL3} we present our new findings on the disorder lines of the model and give the results on the correlation functions
and other thermodynamic parameters in all regions of the phase diagram. The factorized disentangled ground states of the model occurring
on a subset of the disorder lines, are analyzed in Sec.~\ref{Factor}. The regions on the phase diagram where the zero-energy edge Majoranas
exist and their wave functions are presented in Sec.~\ref{MES}. The results are summarized in the concluding Sec.~\ref{Concl}.
%
%
%

%
%
%
\section{Model, spectrum and its zeros}\label{Model}
%
%
%
%

In the spin representation the model is the dimerized quantum $XY$ chain in the presence
of uniform ($h$) and alternating ($h_a$) transverse magnetic fields with the Hamiltonian:
\begin{widetext}
\begin{equation}
\label{XYHam}
   H =- \sum_{n=1}^{N}~   J \Big[ (1+\gamma+\delta (-1)^{n} )
 S_{n}^{x }S_{n+1}^{x}
 + (1-\gamma+ \delta (-1)^{n}) S_{n}^{y} S_{n+1}^{y} \Big]
 +  \big[ h+(-1)^{n} h_a \big] S_{n}^{z} ~.
\end{equation}
\end{widetext}
Here the spin operators $S^\alpha_n= \frac12 \sigma^\alpha_n$ are expressed via the standard Pauli matrices, coupling $J>0$ is ferromagnetic. We assume in our formulas
$0 \leq \delta \leq 1$ and $\gamma \geq 0$. The range of negative $\gamma$ is readily available under exchange
$\gamma \leftrightarrow -\gamma$ and $x \leftrightarrow y$.
The Jordan-Wigner (JW) transformation \cite{LiebSM:1961,Franchini:2017} maps \eqref{XYHam} onto the free
Hamiltonian of spinless fermions
\begin{widetext}
\begin{equation}
\label{XYFermi}
   H =- \sum_{n=1}^{N}~  \frac{J}{2}  \Big[ (1+\delta (-1)^{n} )(c_{n}^\dag c_{n+1} +\mathrm{h.c.})+
                                          \gamma (c_{n}^\dag c_{n+1}^\dag +\mathrm{h.c.}) \Big]
             +  \big( h+(-1)^{n} h_a \big) \big( c_{n}^\dag c_{n}- \frac12~ \big)  ~,
\end{equation}
\end{widetext}
sometimes called in recent literature the (modulated) Kitaev chain \cite{Kitaev:2001}. In the fermionic representation \eqref{XYFermi} the
chain has dimerized hopping and modulated chemical potential. So, whether we deal with spins or with fermions, is a matter of convention.
For the comprehensive analysis of the present model at zero temperature, including its phase diagram, local and non-local order parameters for each phase, we refer readers to the recent work \cite{Chitov:2019}. In this paper we follow the notations of \cite{Chitov:2019}.

The model can be brought to the Bogoliubov-de Gennes spinor form with the $4\times 4$ Hamiltonian  matrix (we set $J=1$):
\begin{equation}
\label{Hk}
  \hat{\mathcal{H}}(k) = \left(%
\begin{array}{cc}
  \hat{A} & \hat{B} \\
  \hat{B}^\dag  & -\hat{A} \\
\end{array}%
\right)~,
\end{equation}
where
\begin{equation}
\label{A}
  \hat{A} \equiv  \left(%
\begin{array}{cc}
  h +\cos k & h_a+i \delta \sin k \\
  h_a-i \delta \sin k & h -\cos k \\
\end{array}%
\right)~,
\end{equation}
and
\begin{equation}
\label{B}
 \hat{B} \equiv  \left(%
\begin{array}{cc}
  -i \gamma \sin k & 0 \\
  0 & i \gamma \sin k \\
\end{array}%
\right)~,
\end{equation}
With a unitary transformation, the Hamiltonian \eqref{Hk} can be brought to the block off-diagonal form
\begin{equation}
\label{HD}
  \hat{\mathcal{H}}^\prime(k)=\left(%
\begin{array}{cc}
  0 & \hat{D}(k) \\
  \hat{D}^\dag (k) & 0 \\
\end{array}%
\right)
\end{equation}
used in the following. Here we defined the operator
\begin{equation}
\label{Dk}
    \hat{D}(k) \equiv \hat{A} (k)+\hat{B}(k)~,
\end{equation}
which has two eigenvalues
\begin{eqnarray}
  \Lambda_\pm (k)&=& h \pm  \frac{1}{\sqrt{2}} \Big(1+2 h_a^2+\delta^2-\gamma^2\nonumber  \\
    &+&  (1-\delta^2+\gamma^2) \cos 2k -2 i \gamma \sin 2k \Big)^{1/2}~.
\label{lambdapm}
\end{eqnarray}
The spectrum of this Hamiltonian consists of four eigenvalues  $\pm E_{\pm}$ \cite{Perk:1975}:
\begin{equation}
\label{Epm}
 E_{\pm}(k)= \frac{1}{\sqrt{2}} \Big| \sqrt{\mathfrak{C}_2(k)+ | \Lambda_+ \Lambda_-|} \pm
                                \sqrt{\mathfrak{C}_2(k)- | \Lambda_+ \Lambda_-|} \Big|~,
\end{equation}
with
\begin{equation}
  \label{C2}
  \mathfrak{C}_2(k) \equiv h^2+h_a^2+\cos^2k+(\delta^2+\gamma^2)\sin^2k~.
\end{equation}

The eigenvalues of $\hat{\mathcal{H}}$ and $\hat{D}$ are related by useful formula
\begin{equation}
\label{C4Lpm}
  E_+ E_- =| \Lambda_+ \Lambda_-|~.
\end{equation}

To extend the earlier analysis \cite{Chitov:2019} of the model to incorporate within the same framework the disorder lines (DLs) and the ground-state
factorization \cite{Chitov:2021}, we perform analytical continuation of the wave numbers onto the complex plane as $(e^{i k})^2 \equiv z$. Then the
first Brillouin zone (BZ) $k \in [-\pi/2, \pi/2]$ maps onto unit circle $|z|=1$.
Eq.~\eqref{C4Lpm} allows us to relate the zeros of the spectra of $\hat{\mathcal{H}}$ and $\hat{D}$ as
\begin{equation}
\label{EpmDD}
E_+^2(z) E_-^2(z) =\mathrm{det}\hat{D}(z) \mathrm{det}\hat{D}^\dag (z)~,
\end{equation}
while the zeros of the spectra of $\hat{A} \pm \hat{B}$ are readily found yielding
\begin{eqnarray}
\label{detD*}
  \mathrm{det}\hat{D}^\dag (z) &=& -\frac{(1+\gamma)^2-\delta^2}{4z}(z-z_+)(z-z_-), \\
   \mathrm{det}\hat{D} (z) &=& -\frac{(1-\gamma)^2-\delta^2}{4z}(z-z_+^{-1})(z-z_-^{-1}).
\label{detD}
\end{eqnarray}
Note that $\hat{D} \leftrightarrow \hat{D}^\dag$  when $\gamma \leftrightarrow -\gamma$, and the zeros of determinants of those matrices are mutually reciprocal:
\begin{equation}
\label{zinv}
  z_\pm(-\gamma)=z_\mp^{-1}~,
\end{equation}
with their explicit expressions given by
\begin{equation}
\label{zpm}
  z_\pm= \frac{\big[ (h^2-h_-^2)^{1/2} \pm (h^2-h_+^2)^{1/2} \big]^2 }{(1+\gamma)^2-\delta^2}
  \equiv e^{2ik_\pm}~,
\end{equation}
and
\begin{equation}
\label{hpm}
  h_\pm^2 \equiv  h_a^2+ \frac12 \Big[ 1+\delta^2-\gamma^2 \pm \big[(1+\gamma)^2-\delta^2 \big]^{1/2}
  \big[(1-\gamma)^2-\delta^2 \big]^{1/2} \Big]
\end{equation}
A word of caution: the above notation $h_\pm^2$ is introduced for further convenience only, and in general  $h_\pm^2 \in \mathbb{C}$.

Quantum phase transitions of the model \eqref{XYHam} correspond to the case when roots \eqref{zpm} lie on the unit circle $|z_\pm|=1$.
For reader's convenience we recapitulate below the properties of the model's phase diagram analysed in detail in \cite{Chitov:2019}:

\textit{(i)} At $|h| > \sqrt{h_a^2+1}$
the model is in the polarized or paramagnetic (PM) phase. At its boundary, shown as bold blue solid lines in Figs.~\ref{HaGam},\ref{DelGam},\ref{HaDelGam}, the root $z_+=1$, and the gap in the spectrum vanishes at the center of the BZ $k=0$.

\textit{(ii)}
At $|h| < \sqrt{h_a^2+1}$ the model is in the ferromagnetic (FM)
phase with the spontaneous longitudinal magnetization $m_x \neq 0$ at $\gamma>0$ and $m_y \neq 0$ at $\gamma<0$.

\textit{(iii)}
The circle on the ($h,\gamma$) plane with the radius $R_\circ =\sqrt{h_a^2+\delta^2}$, shown as bold brown solid line in Figs.~\ref{HaGam},\ref{DelGam},\ref{HaDelGam},
corresponds to the boundary of the topological phase $\mathcal{O}_z(\pi/2)$ with the oscillating string order. At this critical line one of the roots $z_\pm=-1$,
and the gap vanishes at the edge of the BZ $k= \pm \pi/2$.

\textit{(iv)}
Two line segments at $\gamma=0$, shown as bold magenta solid lines in Figs.~\ref{HaGam},\ref{DelGam},\ref{HaDelGam},
correspond to the incommensurate (IC) gapless phase with the complex conjugate roots $|z_\pm|=1$, and the gap vanishing at the IC
wavevector
\begin{equation}
 \label{qIC}
  q= \pm \arcsin \sqrt{\frac{1+h_a^2-h^2}{1-\delta^2}}~.
\end{equation}
This wavevector \eqref{qIC} varies continuously from $q=0$ at the intersection of $\gamma=0$ and $h=\pm\sqrt{h_a^2+1}$
to $q= \pm \pi/2$ where the critical segments end at the intersections with the boundary of the topological phase $\mathcal{O}_z(\pi/2)$ .

The roots  \eqref{zpm} contain also the information about another class of solutions, corresponding to DLs
\cite{Stephenson-I:1970,*Stephenson-II:1970,Stephenson:1970PRB}. These lines, shown in dashed green in Figs.~\ref{HaGam},\ref{DelGam},\ref{HaDelGam},
correspond to the points where the roots $z_\pm$ acquire imaginary parts and become complex conjugate \cite{Chitov:2017PRE,Chitov:2021,Jones:2019}.
See Figs.~\ref{ZpmA} and \ref{ZpmB}.
The regions of the phase diagram bounded by the DLs correspond to the so-called oscillating phases (regimes), where the monotonous exponential decay of the correlation functions is modulated by IC oscillations. In the present model the oscillating regions are localized within the FM phase and smoothly connected to the IC gapless (critical) lines. In addition we found the disorder lines of the second kind when the regime changes abruptly from no oscillations ($q=0$) to the commensurate oscillations ($q=\pi/2$). Our new results presenting the detailed analysis of the model's DLs for three special cases are given in Sec.~\ref{DL3}.

%
%
%
\section{Disorder lines}\label{DL3}
%
%
%
%
%
%
\subsection{Special case: $h_a \neq 0$ and $\delta=0$}\label{HaGamSec}
\begin{figure}[]
\centering{\includegraphics[width=8.0cm]{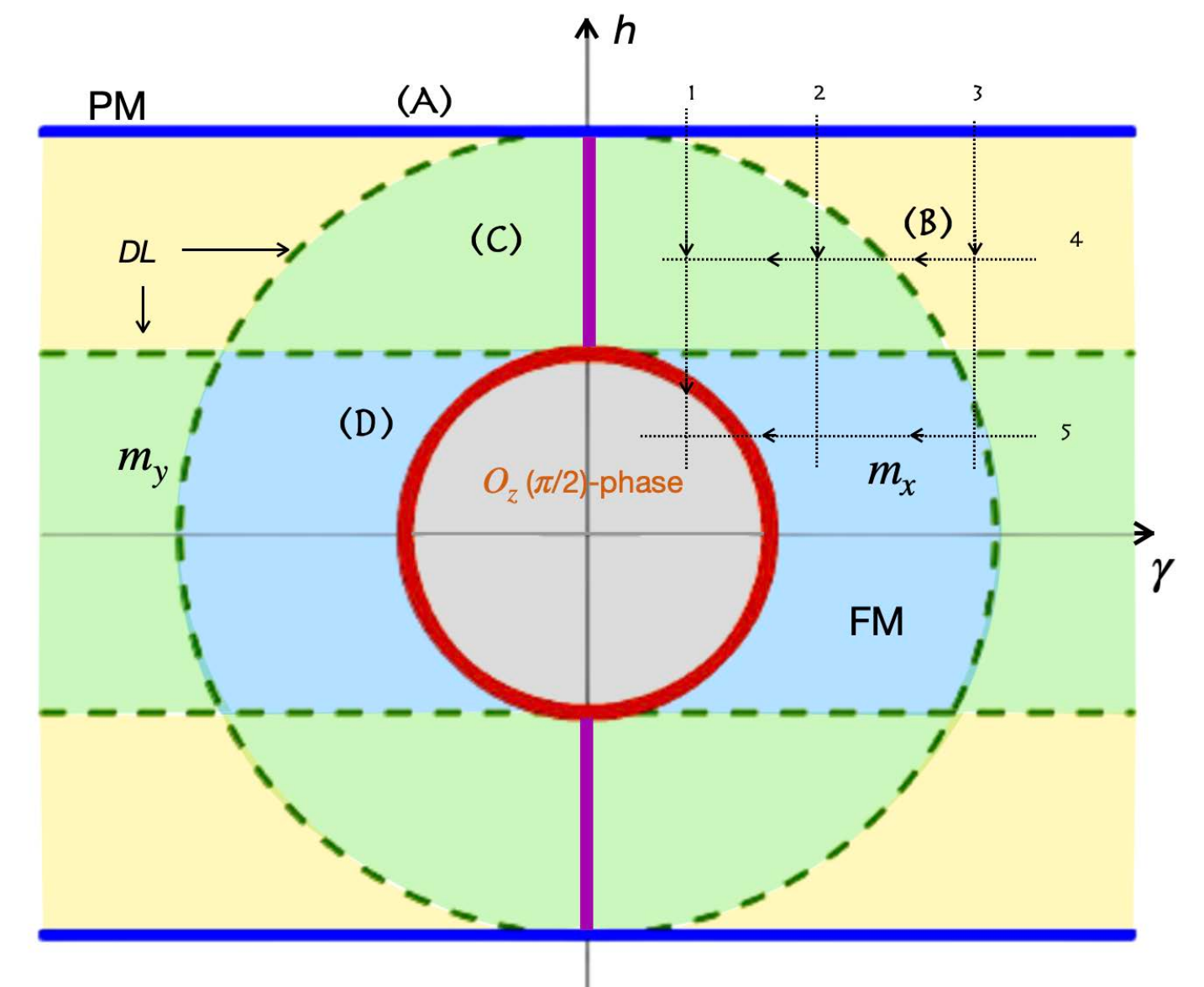}}
\caption{The ground state phase diagram on ($h,\gamma$) plane for the case of zero dimerization $\delta=0$ and non-zero staggered field $h_a \neq 0$. The bold solid lines denote the phase boundaries: PM-FM in blue; FM-$\mathcal{O}_z$ in brown; gapless IC in magenta. The disorder lines are localized within the ferromagnetically-ordered phase and shown in dashed green. The regions with IC oscillations, labeled by (C) are shown in light green; (B) regions shown in light yellow contain no oscillations since $q=0$; in regions labeled by (D) and shown in light blue, the oscillations are commensurate with $q=\pi/2$.}
\label{HaGam}
\end{figure}

\begin{figure}[]
\centering{\includegraphics[width=7.5cm]{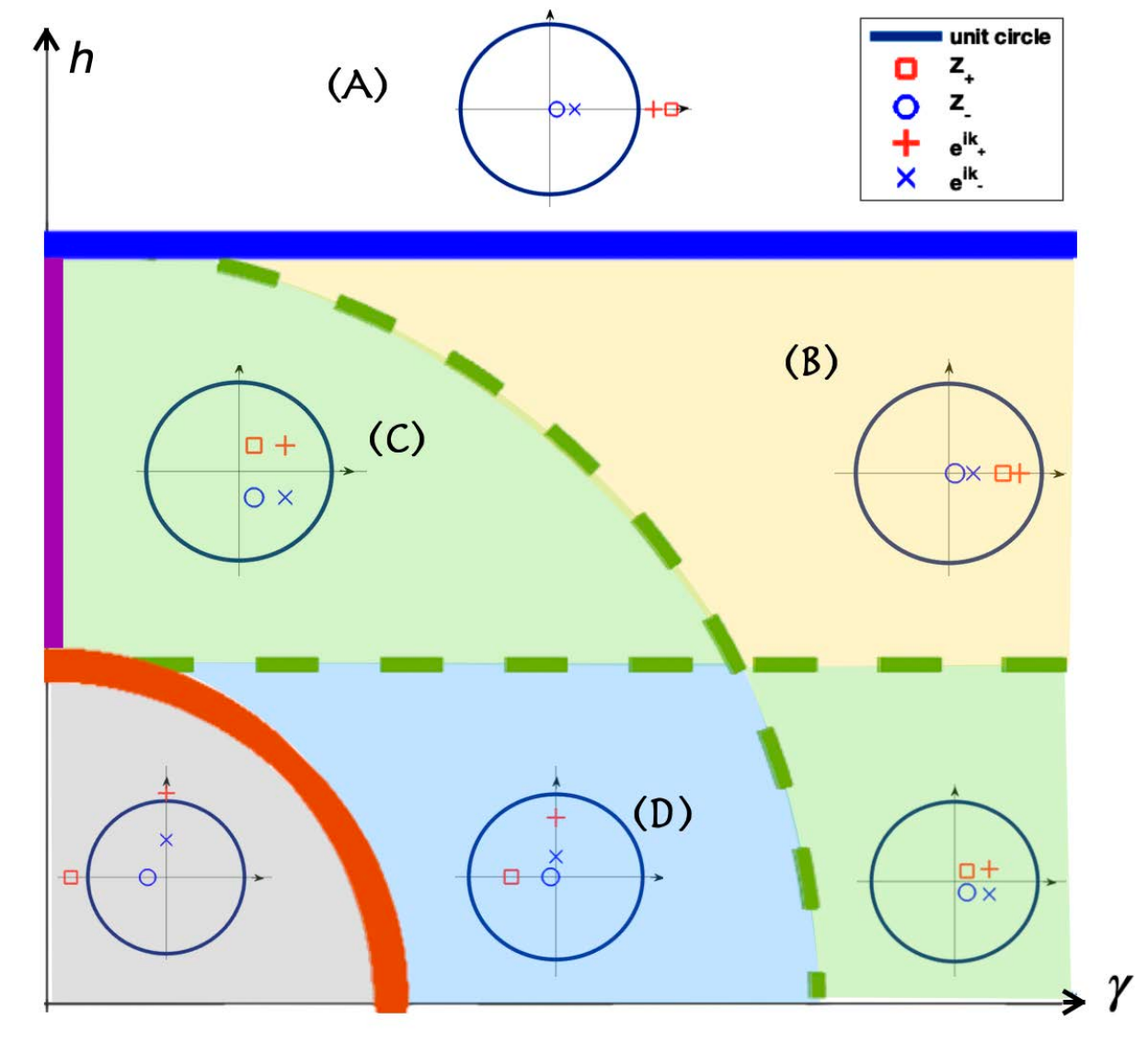}}
\caption{The roots $z_\pm$ and $e^{ik_\pm}$  ($z_\pm \equiv e^{i2k_\pm}$) determined by Eqs. \eqref{zpm} and \eqref{zpmHa} for $\delta=0$ and $h_a \neq 0$ in the first quadrant of the phase diagram (see Fig.~\ref{HaGam}). The positions of the roots are shown in the complex plane with respect to the unit circle $|z|=1$ drawn by solid blue line. }
\label{ZpmA}
\end{figure}

\begin{figure}[]
\centering{\includegraphics[width=9.0cm]{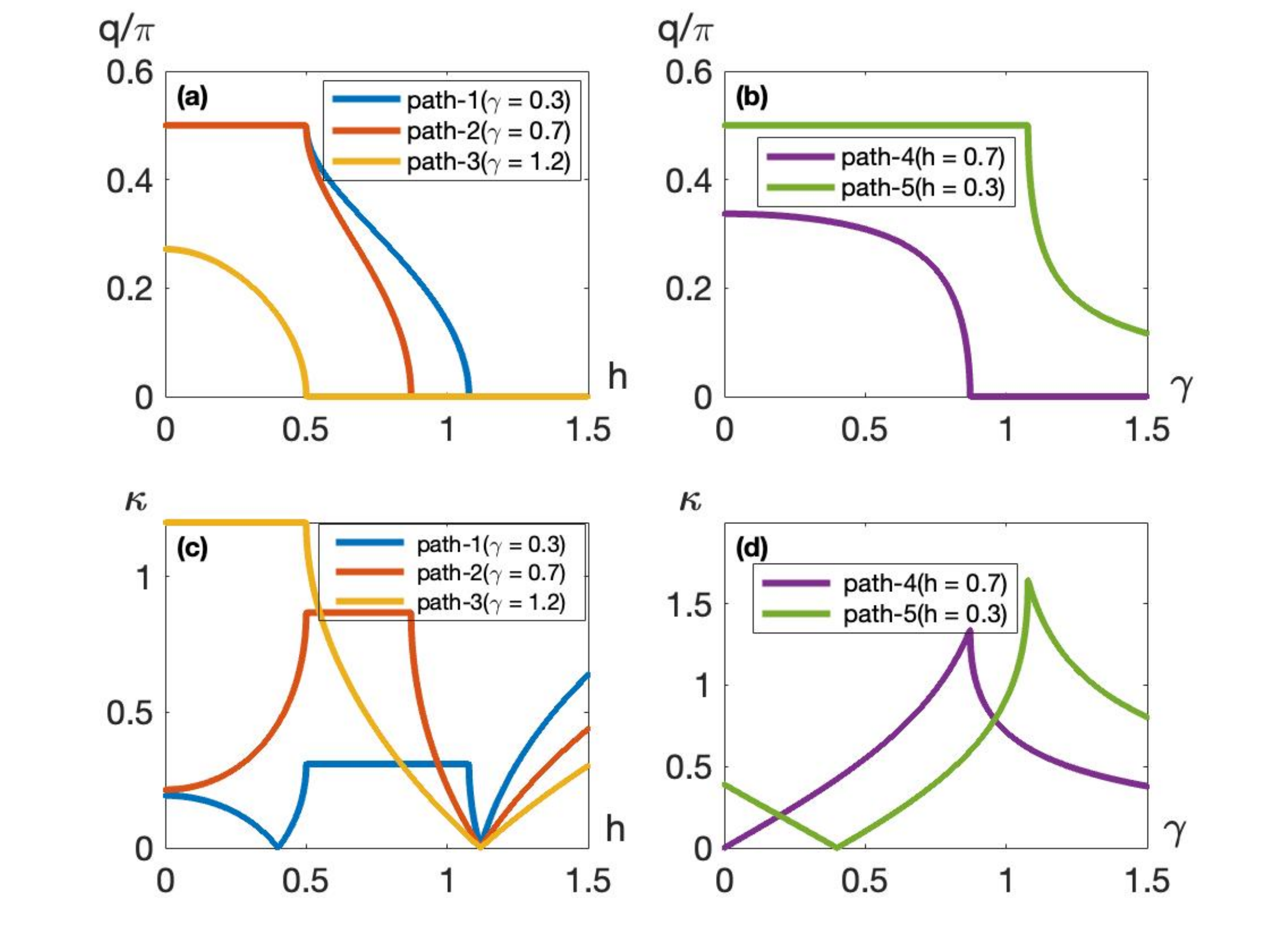}}
\caption{The modulation wave number ($q$) and the inverse correlation length ($\kappa$)  along several paths indicated in ($h,\gamma$) plane,  see Fig.~\ref{HaGam} for the case $\delta=0$ and $h_a \neq 0$. The disorder lines bounding the regions of the IC wave numbers $q$ are accompanied by the cusps of $\kappa$, while the quantum phase transitions PM-FM and FM-$\mathcal{O}_z$ correspond to $\kappa=0$.}
\label{KapQA}
\end{figure}

%
%
The boundaries of the oscillating regions \eqref{hpm} (DLs) are quite simple in this case:
\begin{equation}
\label{hpmHa}
 h^2=h^2_\pm =
\left\{
\begin{array}{lr}
1+h_a^2-\gamma^2 \\[0.2cm]
h_a^2  \\
\end{array}
\right.
\end{equation}
The roots \eqref{zpm} are given by:
\begin{equation}
\label{zpmHa}
  z_\pm= \Big[ \frac{(h^2-h_a^2)^{1/2}  \pm (h^2+\gamma^2-1-h_a^2)^{1/2}   }{1+\gamma} \Big]^2~.
\end{equation}
The phase diagram including the DLs for this case is depicted in Fig.~\ref{HaGam}. The complex-valued
roots $z_\pm$ are localized between intersections of the ``disorder circle" of radius $R_{\s DL}=\sqrt{1+h_a^2}$ and the lines
$h = \pm h_a$. They are shown in dashed green in Fig.~\ref{HaGam}. The radius of the topological circle $R_\circ=h_a$.
Note that in the limit $h_a \to 0$ one recovers the results of Barouch and McCoy \cite{McCoyII:1971} with a single DL
determined by the circle $h^2+\gamma^2=1$.
With parametrisation of the roots \eqref{zpm} as $z(k)$, where $k \in \mathbb{C}$ and
\begin{equation}
\label{kC}
 k \equiv q+i \kappa, ~z=e^{2(iq -\kappa)}~,
\end{equation}
$q$ yields the wave number of oscillations, while $\kappa$ is the inverse correlation length \cite{Chitov:2017PRE,Chitov:2021}.
Evolution of the roots \eqref{zpmHa} on the complex plane and dependencies of $q$ and $\kappa$ along several paths on the ($h,\gamma$) plane (see Fig.~\ref{HaGam}) are presented in Figs.~\ref{ZpmA} and \ref{KapQA}. As expected from general arguments, the lines
of quantum criticality correspond to the roots $z_\pm$ \eqref{zpmHa} lying on the unit circle on the complex plane and vanishing $\kappa$. DLs are weaker singularities and they are accompanied by cusps in the correlation length.


The IC oscillations with complex conjugate roots $z_\pm$ are localized in the regions colored in green in Figs.~\ref{HaGam} and \ref{ZpmA}. For those regions in the first quadrant one can find explicitly:
\begin{equation}
\label{qICHa}
 q=
\left\{
\begin{array}{lr}
 \arcsin \sqrt{\frac{1+h_a^2-h^2-\gamma^2}{1-\gamma^2}},~[h>h_a] \cup [h^2+\gamma^2< R_{\s DL}^2 ] \\[0.2cm]
 \arcsin \sqrt{\frac{h_a^2-h^2}{\gamma^2-1}},~[h<h_a] \cup [h^2+\gamma^2> R_{\s DL}^2 ] \\
\end{array}
\right.
\end{equation}
and
\begin{equation}
\label{kapHa}
  \kappa = -\frac12 \ln \frac{|\gamma-1|}{\gamma+1}~.
\end{equation}
The DLs found for this case (see Fig.~\ref{HaGam}) are of the first kind, since the wave number of the IC oscillations changes
continuously between the boundaries of commensurate regions, see Fig.~\ref{KapQA}.
Outside of the IC regions (FM phase) the wave number is:
\begin{equation}
\label{qHa}
 q=
\left\{
\begin{array}{lr}
 0~,~~[h>h_a] \cup [h^2+\gamma^2> R_{\s DL}^2 ] \\[0.2cm]
 \pi/2~,~~[h<h_a] \cup [h^2+\gamma^2< R_{\s DL}^2 ]  \\
\end{array}
\right.
\end{equation}
These two regions are labelled as B ($q=0$) and D ($q=\pi/2$) in Figs.~\ref{HaGam} and \ref{ZpmA}.

\begin{figure*}[t]
\centering{\includegraphics[width=\textwidth]{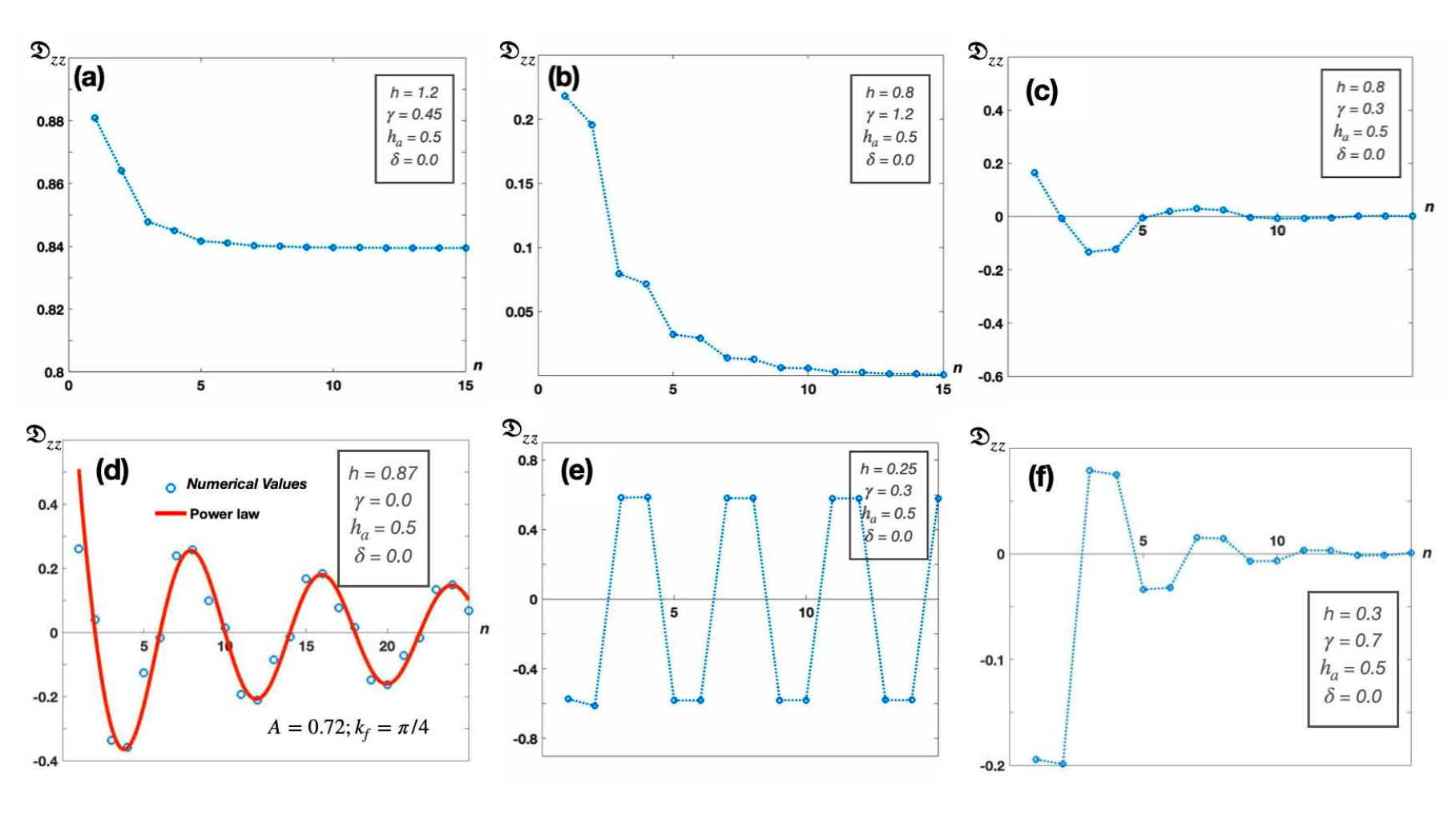}}
\caption{The characteristic behavior of the string-string correlation function $\mathfrak{D}_{zz}$ in different phases and regimes of oscillations
shown in the phase diagram, see Fig.~\ref{HaGam}. The panels (a), (d), and (e) show the correlation functions in the PM, gapless IC, and oscillating string-ordered $\mathcal{O}_z(\pi/2)$, respectively. The panels (b), (c), and (f) correspond to three regions of the FM phase: (b) $q=0$, no oscillations;
(c) gapped IC oscillations; (f) $q=\pi/2$ oscillations. More explanations are given in the text.}
\label{CFA}
\end{figure*}

The results for the correlation functions in different phases and regimes of oscillations are collected in Fig.~\ref{CFA}.
It has been checked by explicit calculations that the types of behaviors are qualitatively equivalent for all spin-spin and
string-string correlation functions. In Fig.~\ref{CFA} we plot the representative numerical results for the string correlation function
\begin{equation}
\label{Dzz}
  \mathfrak{D}_{zz}(L,R) \equiv
 \Big \langle \prod_{l=L}^{R} \sigma_l^z \Big \rangle =
 \Big \langle  \prod_{l=L}^{R} \big[ i b_l a_l \big]  \Big \rangle~.
\end{equation}
$\mathfrak{D}_{zz}$ is defined in terms of the original spin operators $\sigma_l^z$
or Majorana fermions related to the JW fermions as
\begin{equation}
\label{Maj}
   a_n +i b_n  \equiv 2 c^{\dag}_n~.
\end{equation}
The correlation function \eqref{Dzz} is calculated numerically from the determinant of the block Toeplitz matrix whose elements, i.e., the two-point
Majorana correlators  $\langle ib_na_m \rangle$ are explicitly given in \cite{Chitov:2019}.

As one can see from Fig.~\ref{CFA}, the behavior of the correlation function is easily understood from the structure of the complex roots depicted in
Figs.~\ref{ZpmA} and \ref{KapQA}. The plot (a) corresponds to the polarized phase (PM) where $\mathfrak{D}_{zz}(1,n)$ monotonously decays to the
limiting value $\mathcal{O}_{z}^2$, where $\mathcal{O}_{z}$ can be identified as a single string order parameter of the PM phase \cite{Chitov:2022}.

In the panel (b) of Fig.~\ref{CFA} the string correlation function is shown for the region (B) on the phase diagram in Fig.~\ref{HaGam} where oscillations are absent.
This is the ferromagnetic phase where the order parameter is the spontaneous longitudinal magnetization $m_x$, while the string order is absent, and
accordingly, $\mathfrak{D}_{zz}$ decays to zero \cite{Chitov:2019}.  The exponential decay of $\mathfrak{D}_{zz}$ modulated by the IC oscillations, shown in
panel (c) corresponds to the IC region (C) of the FM phase, see Fig.~\ref{HaGam}.

The IC gapless phase shown in magenta in Fig.~\ref{HaGam} is algebraically ordered, $\mathfrak{D}_{zz}$ demonstrates the power-law decaying oscillating behavior:\cite{Chitov:2020}
\begin{equation}
\label{DzzIC0}
  \mathfrak{D}_{zz}(1,n) = \frac{\mathcal{A}}{\sqrt{n}} \cos(k_{\s F} n)~.
\end{equation}
The wave number of oscillations in this phase is also the Fermi wave number of the JW fermions $q= k_{\s F}$, given by
Eq.~\eqref{qIC}. The explicit formula for the amplitude $\mathcal{A}$ is unavailable. For a particular choice of parameters yielding $k_{\s F}=\pi/4$, the results are shown in Fig.~\ref{CFA}(d) with the fit $\mathcal{A} \approx 0.72$. We found an excellent agreement between Eq.~(\ref{DzzIC0}) and direct
numerics.

The panel (e) of Fig.~\ref{CFA} shows the characteristic behavior of $\mathfrak{D}_{zz}$ in the phase labelled as $\mathcal{O}_z(\pi/2)$. The correlation function oscillates with the period of four lattice spacings, and its limiting value yields the string order parameter, see Ref.~\cite{Chitov:2019} for details. In panel (f) we show the correlation function in region (D) of the phase diagram: the four-site oscillations ($q=\pi/2$) are present, but the
string order vanished ($\mathfrak{D}_{zz} \to 0$), since it is the FM phase ($m_x \neq 0$).

%
%
%

%
%
%
\subsection{Special case: $h_a = 0$ and $\delta \neq 0$}\label{DelGamSec}
\begin{figure}[]
\centering{\includegraphics[width=8.5cm]{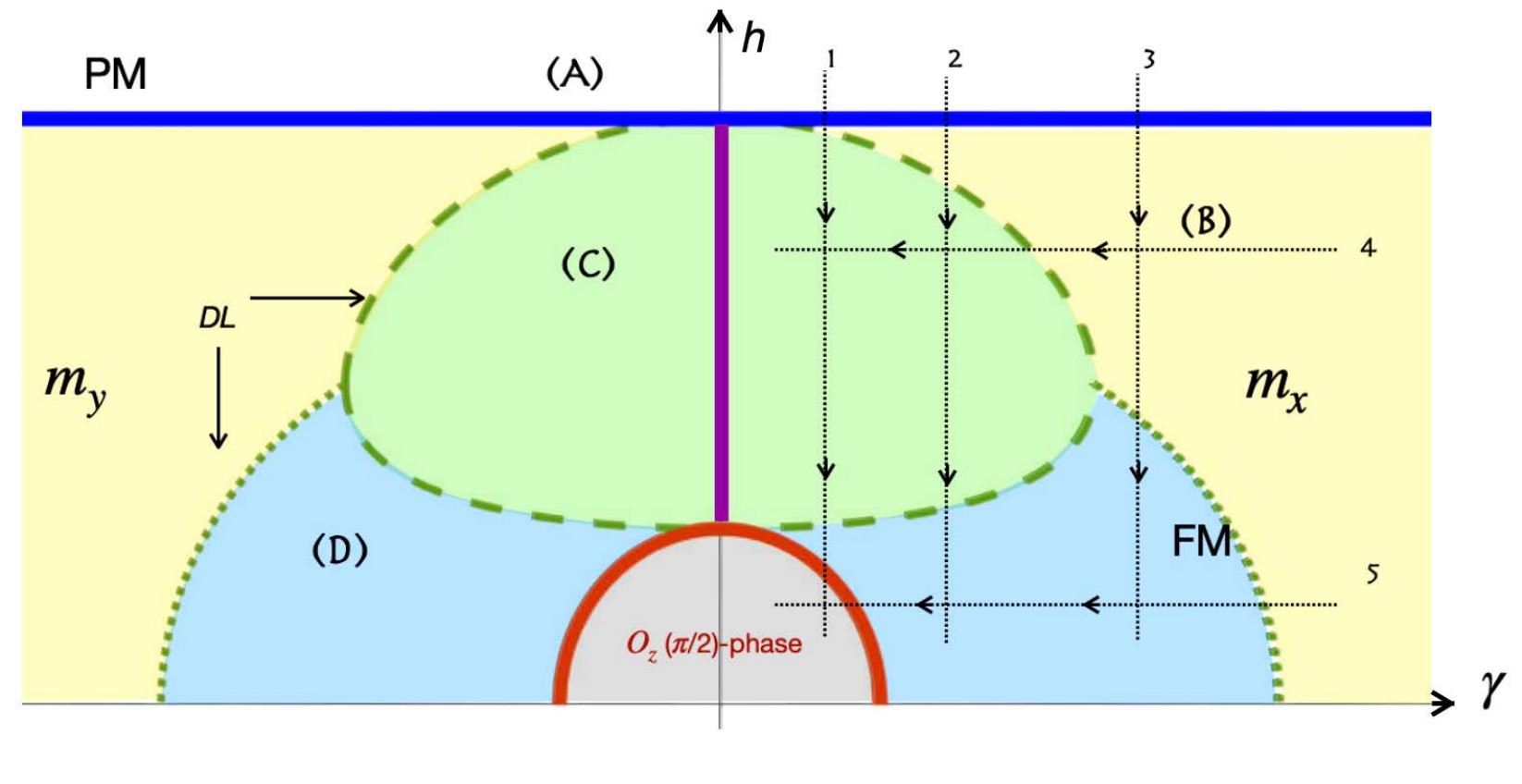}}
\caption{The ground state phase diagram in ($h,\gamma$) plane for the case $\delta \neq 0$ and $h_a = 0$. The bold solid lines denote the phase boundaries: PM-FM in blue; FM-$\mathcal{O}_z$ in brown; gapless IC in magenta. The disorder lines are localized within the FM phase. The disorder lines of the first kind determined by the real solutions of \eqref{hpmDel} are shown in dashed green. They bound the region with IC oscillations (C) shown in green.  (B) regions with no oscillations ($q=0$) shown in light yellow; (D) regions with the commensurate oscillations ($q=\pi/2$) shown in light blue. Regions (B) and (D) are separated by the disorder lines of the second kind given by \eqref{DL2expl} and plotted in dotted green.}
\label{DelGam}
\end{figure}

%
%
%
In this case the IC oscillating region shown in green in Fig.~\ref{DelGam} is bounded by two DLs
\begin{equation}
\label{hpmDel}
  h^2=h_\pm^2 =  \frac12 \Big[ 1+\delta^2-\gamma^2 \pm \big[(1+\delta^2-\gamma^2)^2-4 \delta^2 \big]^{1/2} \Big]~,
\end{equation}
depicted by dashed green lines.
In ($h,\gamma$) upper half-plane shown in Fig.~\ref{DelGam}, the DLs (dashed green) are located between the paramagnetic phase $h >1$ and the topological circle of radius $R_\circ=\delta$.
\begin{equation}
\label{hpm0}
 \mathrm{At~} \gamma=0:~h_\pm =
\left\{
\begin{array}{lr}
1 \\[0.2cm]
\delta  \\
\end{array}
\right.
\end{equation}
and two curves $h_\pm(\gamma)$ smoothly intersect ($h_+=h_-$) at $\gamma=\pm(1-\delta)$.

The inverse correlation length inside the IC oscillating region (C),  shown by green, is
\begin{equation}
\label{kapDel}
  \kappa = -\frac14 \ln \frac{(1-\gamma)^2-\delta^2}{(1+\gamma)^2-\delta^2}~,
\end{equation}
while the wave number of oscillations grows continuously from $q=0$ on the upper boundary of the IC region, to $q=\pi/2$ on its lower boundary,
following the formula:
\begin{equation}
 \label{qDel}
  q= \arcsin \sqrt{\frac{h_+^2-h^2}{h_+^2-h_-^2}}~.
\end{equation}
The oscillating IC phase becomes gapless on the critical line $\gamma=0$ shown by bold magenta on the phase diagram in Fig.~\ref{DelGam}.

\begin{figure}[]
\centering{\includegraphics[width=7.0cm]{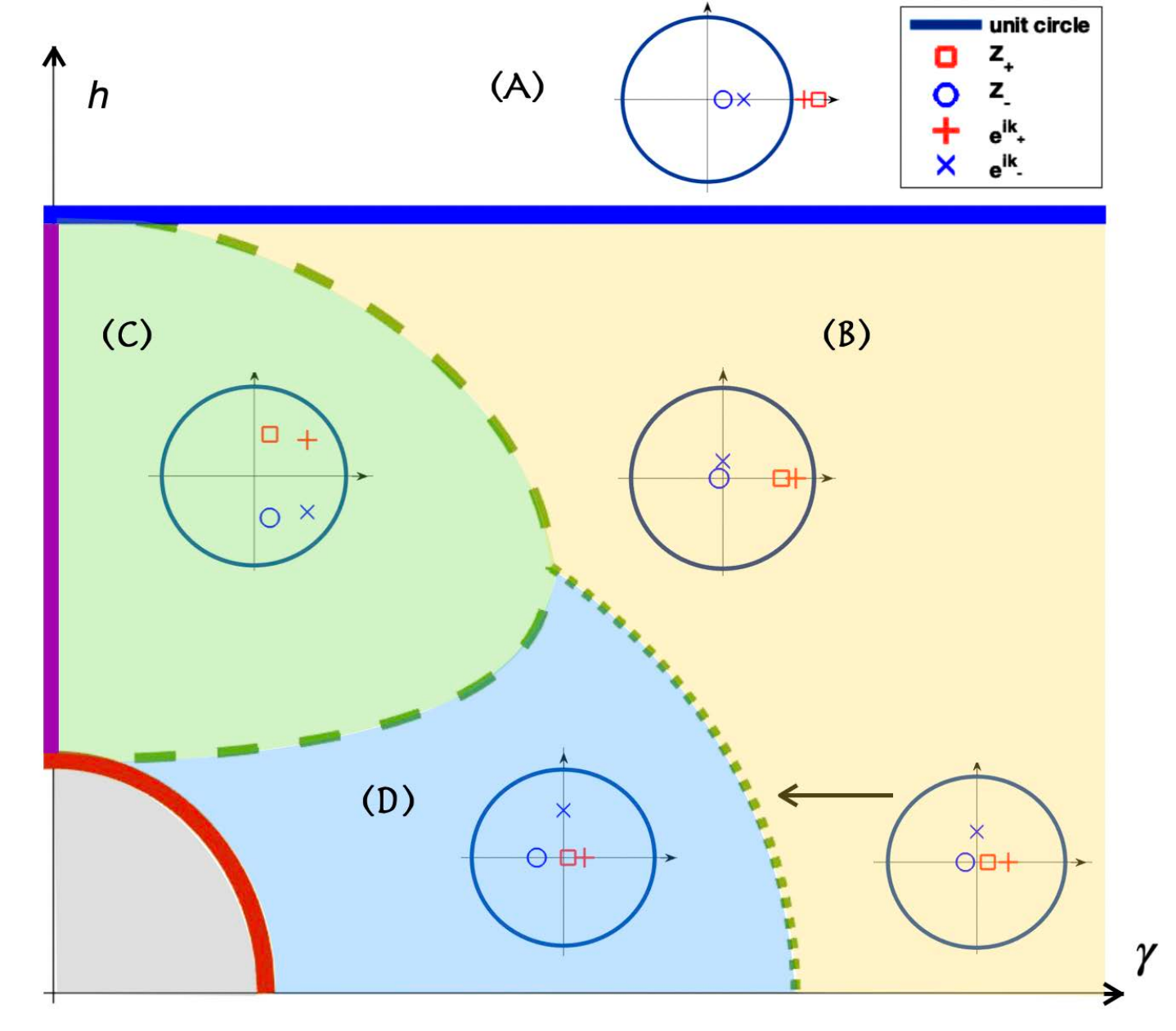}}
\caption{The roots $z_\pm$ and $e^{ik_\pm}$  ($z_\pm \equiv e^{i2k_\pm}$) determined by Eqs. \eqref{zpm} and \eqref{hpmDel} for $\delta \neq 0$ and $h_a =0$ in the first quadrant of the phase diagram (see Fig.~\ref{DelGam}), are shown in the complex plane with respect to the unit circle $|z|=1$ drawn by solid blue line. Two roots on the DL2
\eqref{DL2expl} when the condition \eqref{DL2} is satisfied are drawn for a particular point $\gamma=1.005$, $\delta =0.3$: $|e^{i k_\pm}|=0.389$.}
\label{ZpmB}
\end{figure}
\begin{figure}[]
\centering{\includegraphics[width=9.0cm]{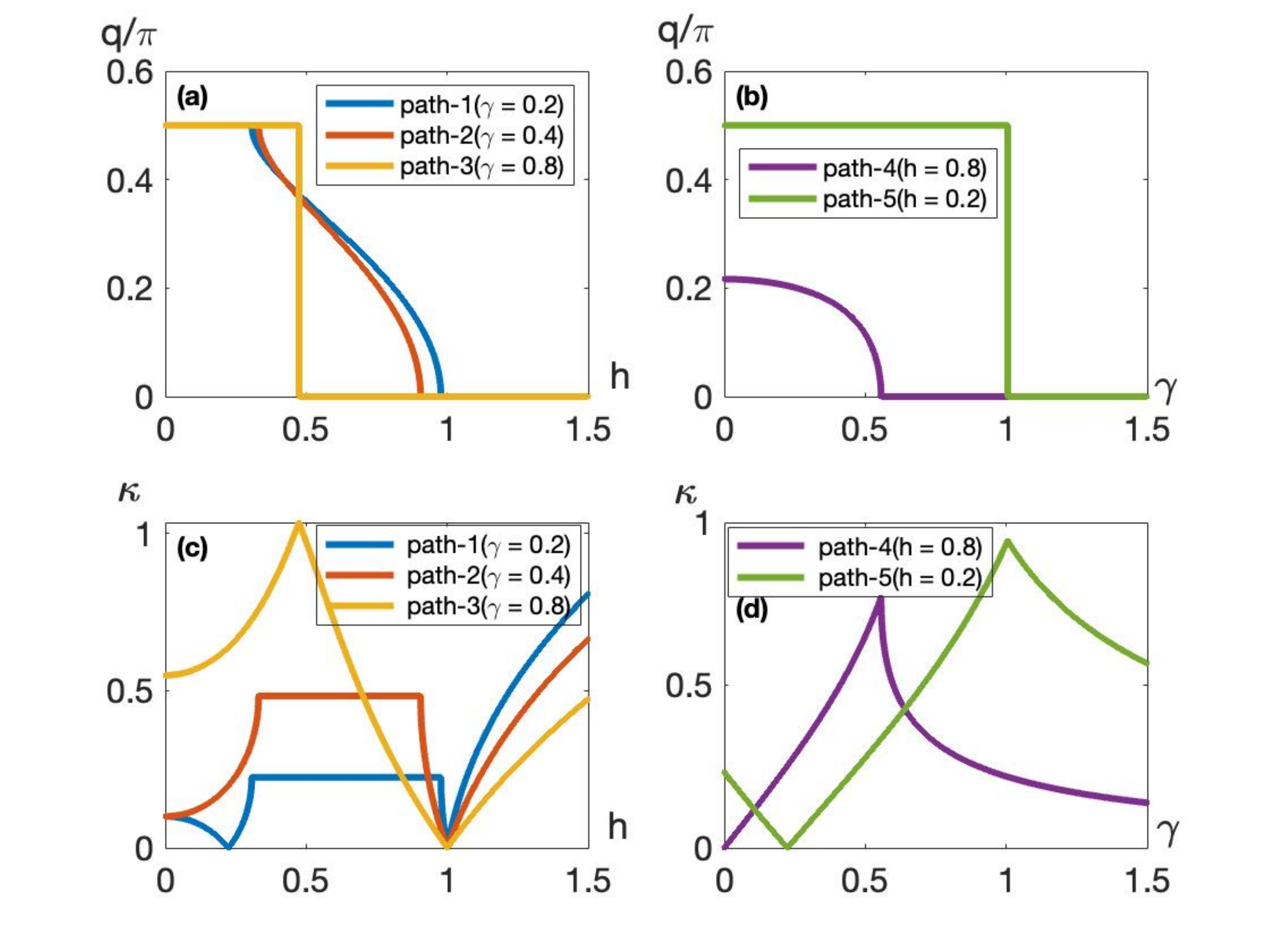}}
\caption{The modulation wave number ($q$) and the inverse correlation length ($\kappa$)  along several paths indicated in ($h,\gamma$) plane,  see Fig.~\ref{DelGam} for the case $\delta \neq 0$ and $h_a = 0$. The disorder lines are accompanied by the cusps of $\kappa$, while the quantum phase transitions PM-FM and FM-$\mathcal{O}_z$ correspond to $\kappa=0$. The wave number is continuous, albeit not smooth on the disorder lines of the first kind, while it undergoes a discontinuity on the lines of the second kind.}
\label{KapQB}
\end{figure}

We have also found the so-called \textit{disorder lines of the second kind} (DL2) \cite{Stephenson:1970PRB}, such that the wave vector of modulations changes discontinuously \cite{Nussinov:2011,Nussinov:2012,Chitov:2021} upon crossing those lines. In the case under consideration the modulation transition with a jump from $q=0$ to $q=\pi/2$
occurs when
\begin{equation}
\label{DL21}
  h_\pm^2 \in \mathbb{C},~~h_+ =h_-^\ast~.
\end{equation}
The DLs of the first kind analysed above, are characterized by continuous appearance of the IC modulation, when the roots \eqref{zpm} expand onto complex values, such that $z_+=z_-^\ast$ and the continuous  evolution of the modulation wave number $q$ follows
the smooth growth of the phase of the complex numbers $z_\pm$, see Figs.~\ref{ZpmB} and \ref{KapQB}.

The mechanism of the discontinuous modulation (DL2) is quite different: one can check that in the range where $h_\pm^2$ are complex conjugate,  one root of \eqref{zpm} $e^{ik_+}=a$ is real, while the second root $e^{ik_-}=ib$ is purely imaginary ($a,b>0$), and the transition of the regime occurs
when
\begin{equation}
\label{DL2}
|e^{ik_+}|=|e^{ik_-}|~,
\end{equation}
see Fig.~\ref{ZpmB}.
At this point the correlation length demonstrates a cusp  as at other DLs (see Fig.~\ref{KapQB}), while $q$ undergoes an abrupt jump following the phase of the root: from $\Im (e^{ik_+})=0$ to $\Im (e^{ik_-})=\pi/2$, see  Fig.~\ref{KapQB}. One can verify that the condition  \eqref{DL2} amounts to $h^2=\Re (h_\pm^2)$, yielding the equation of the DL2:
\begin{equation}
\label{DL2expl}
h^2=\frac12(1+\delta^2-\gamma^2)~.
\end{equation}
These DL2s on $(h,\gamma)$ plane are shown in Fig.~\ref{DelGam} in dotted green: they start at the point of intersection of $h_+$ and $h_-$ at $h=\sqrt{\delta}$ and $\gamma=\pm(1-\delta)$, and reach $h=0$ at  $\gamma=\pm \sqrt{1+\delta^2}$.

One should keep in mind that the roots solving the equation for zeros of the spectrum enter the analytic expressions for correlation functions on the same footing \cite{Chitov:2021}. The asymptotes of correlators are controlled however by the leading contribution from the root corresponding to minimal $\kappa$ which yields the inverse correlation length. In case of DL2, the  roots $e^{ik_\pm}$ resulting in, respectively, the monotonous/oscillating contributions to asymptotes of correlation functions, are close in absolute values near DL2, cf. Eq.~\eqref{DL2}, so their contributions are comparable, and thus appearance and vanishing of oscillations are smeared across DL2. This analytical conclusion is confirmed by direct numerical calculations of various correlation functions.

The qualitative behavior of the correlation functions is controlled by the properties of the roots in different regions of the phase diagram, see Fig.~\ref{ZpmB}. From comparison of Figs.~\ref{ZpmA} and \ref{ZpmB}, one can see that the only difference in the root structure between the case A
of previous subsection and the current case B, is that for the latter the appearance of the real and imaginary parts of $e^{i k_\pm}$ is not synchronized in the regime without oscillations ($q=0$) and with commensurate oscillations ($q=\pi/2$). However, since the asymptotes of correlation functions are controlled by the root with minimal $\kappa$ (i.e. the closest to the unit circle), then the correlation functions in different regions of the phase diagram have the behavior similar to that shown in Fig.~\ref{CFA}.

One can check that the results of the present subsection recover those of Barouch and McCoy \cite{McCoyII:1971} in the limit $\delta \to 0$.

%
%
%
\subsection{General case: $h_a \neq  0$ and $\delta \neq 0$}\label{HaDelGamSec}
\begin{figure}[]
\centering{\includegraphics[width=8.5cm]{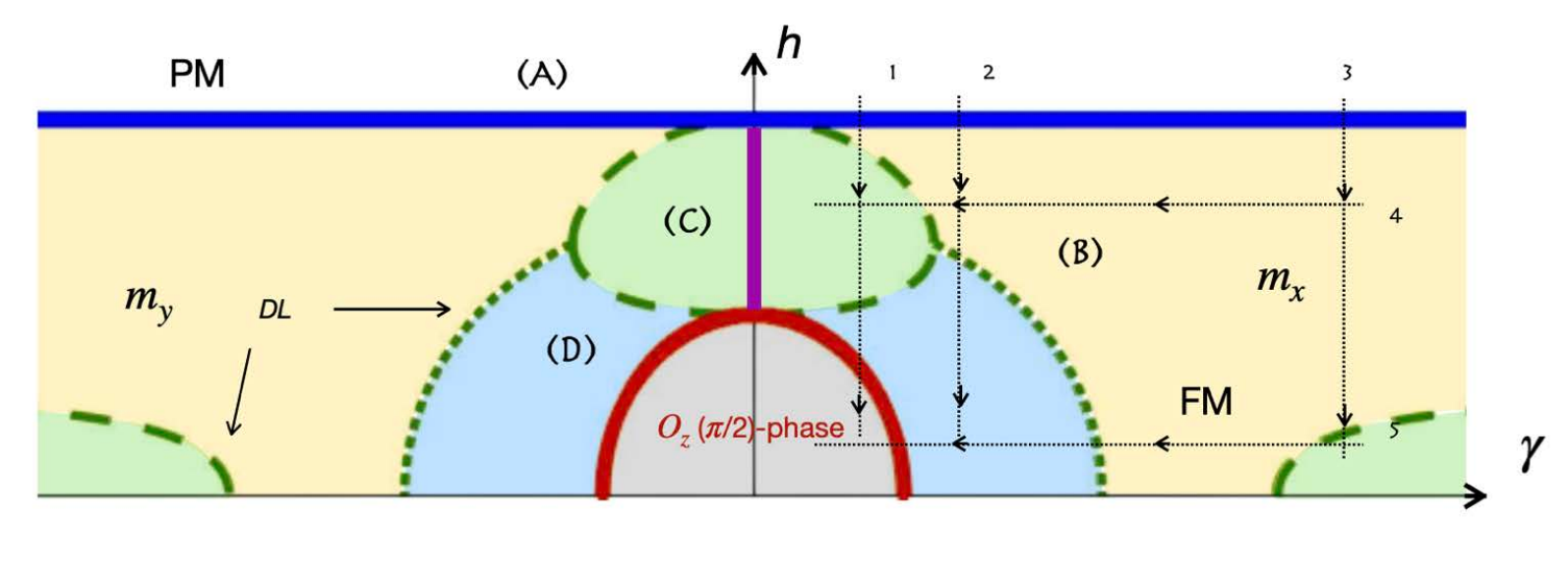}}
\caption{The ground state phase diagram in ($h,\gamma$) plane for the general case $\delta \neq 0$ and $h_a \neq 0$. The bold solid lines denote the phase boundaries: PM-FM in blue; FM-$\mathcal{O}_z$ in brown; gapless IC in magenta. The disorder lines are localized within the FM phase. The disorder lines of the first kind determined by the real solutions of \eqref{hpm} are shown in dashed green. They bound the region with IC oscillations (C) shown in green. (B) regions with no oscillations ($q=0$) shown in light yellow; (D) regions with the commensurate oscillations ($q=\pi/2$) shown in light blue. Regions (B) and (D) are separated by the disorder lines of the second kind given by \eqref{DL2expl2} and plotted in dotted green.}
\label{HaDelGam}
\end{figure}

%
%
%
The ground state phase diagram for the general case is presented in Fig.~\ref{HaDelGam}. Similarly to the previous case with $h_a=0$,
it contains the central IC oscillating region shown in green, bounded by two DLs $h=h_\pm$ ($h_\pm \in \mathbb{R}$,
cf. definitions \eqref{hpm} ) and located between the paramagnetic phase $h >\sqrt{1+h_a^2}$ and the topological circle of radius $R_\circ=\sqrt{\delta^2+h_a^2}$.

In addition, non-zero $h_a$ brings about a possibility of another real solutions for $h_\pm$, engendering two extra IC wings at $|\gamma| \gtrsim 1+\delta$. The detailed shape of those wings and of the DL2 depend slightly on the relations between parameters:

\underline{\textit{(i)} $h_a<\delta$:} In this case the IC oscillating wing is bounded by $h_+ \geq 0$  at  $|\gamma| \geq \gamma_\circ$, where
\begin{equation}
\label{gam0}
  \gamma_\circ \equiv \frac{\sqrt{1+h_a^2}\sqrt{\delta^2+h_a^2}}{h_a}~,
\end{equation}
while  the other branch $h_-$ is not pertinent in this range, since $h_-^2<0$ at  $|\gamma|> 1+\delta$. Note that
\begin{equation}
\label{gams}
 \forall~|\gamma|>1+\delta:~h^2_\pm \in \mathbb{R}~.
\end{equation}
As explained above, the DL2 can occur only in the range where $h^2_\pm$ are complex, which is, according to \eqref{hpm}:
\begin{equation}
\label{HpmCom}
 \forall~ 1-\delta<|\gamma|<1+\delta:~h^2_\pm \in \mathbb{C}~.
\end{equation}
The DL2 equation $h^2=\Re (h_\pm^2)$ yields the parabola
\begin{equation}
\label{DL2expl2}
h^2=h_a^2+ \frac12(1+\delta^2-\gamma^2)~.
\end{equation}
The DL2s shown in dotted green  in Fig.~\ref{HaDelGam}, start at $h=h_+=h_-=\sqrt{h_a^2+\delta}$ and $\gamma=\pm(1-\delta)$ and reach $h=h_+=0$ at  $\gamma=\pm \sqrt{1+\delta^2+2h_a^2}$. This is the case plotted in Fig.~\ref{HaDelGam}.

\underline{\textit{(ii)} $h_a>\delta$:} In this case (not shown) not only $h_+^2  \in \mathbb{R}$, but also $h_+^2>0$ at $|\gamma|>1+\delta$. The other
branch $h_-^2>0$ at $1+\delta<|\gamma|<\gamma_\circ$.  The DL2s span through the whole range of complex-valued $h^2_\pm$, cf. condition \eqref{HpmCom}. These lines are located between the points $h=h_+=h_-=\sqrt{h_a^2+\delta}$ at $\gamma=\pm(1-\delta)$ and $h=h_+=h_-=\sqrt{h_a^2-\delta}$ at $\gamma=\pm(1+\delta)$.

\begin{figure}[]
\centering{\includegraphics[width=7.5cm]{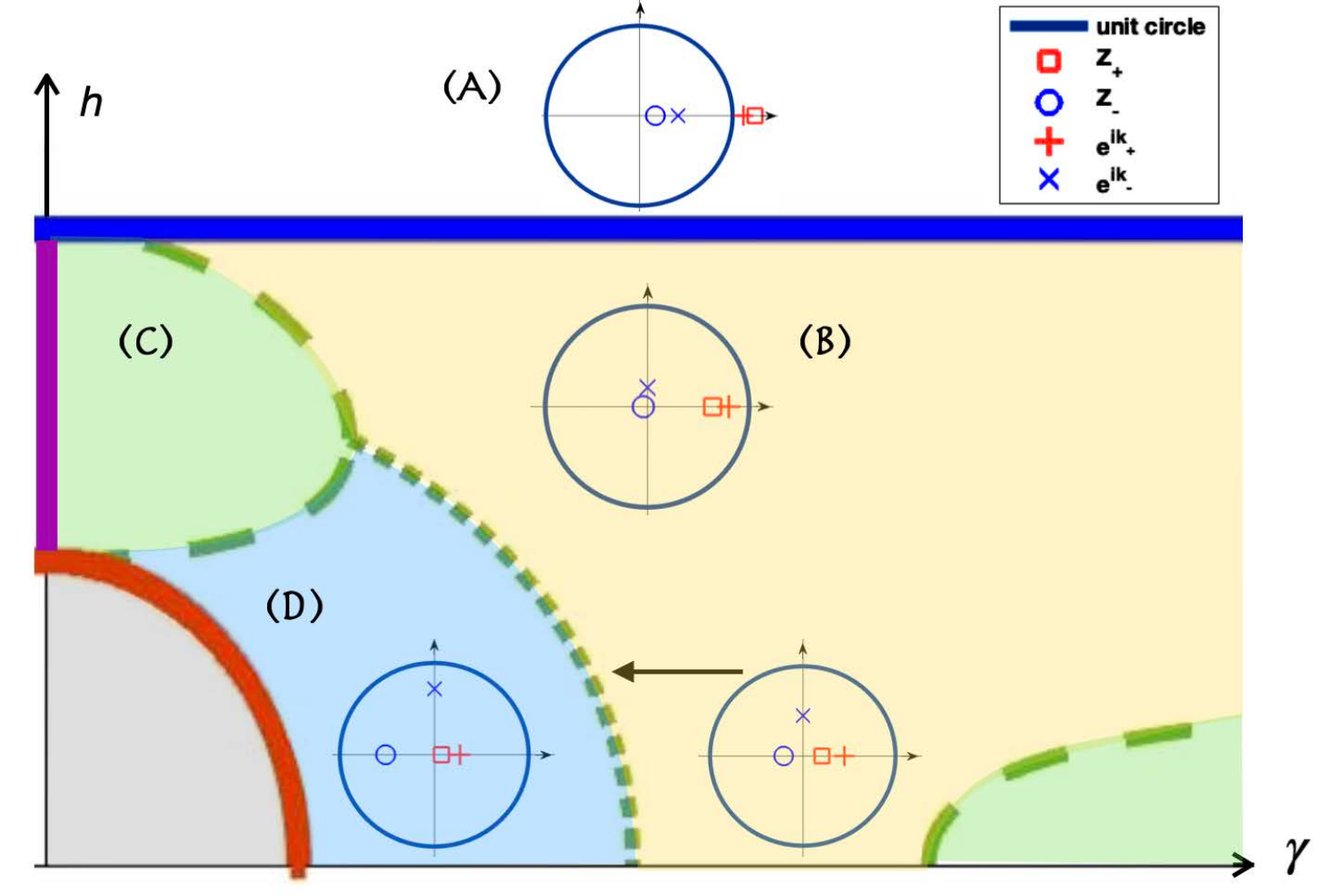}}
\caption{The roots $z_\pm$ and $e^{ik_\pm}$  ($z_\pm \equiv e^{i2k_\pm}$) determined by Eqs. \eqref{zpm} and \eqref{hpm} for $\delta \neq 0$ and $h_a \neq 0$ in the first quadrant of the phase diagram (see Fig.~\ref{HaDelGam}), are shown in the complex plane with respect to the unit circle $|z|=1$ drawn by solid blue line. Two roots on the DL2
\eqref{DL2expl2} when the condition \eqref{DL2} is satisfied are drawn for a particular point $\gamma=1.0$, $\delta =0.3$, $h_a=0.4$: $|e^{i k_\pm}|=0.4518$.}
\label{ZpmC}
\end{figure}

\begin{figure}[]
\centering{\includegraphics[width=9.0cm]{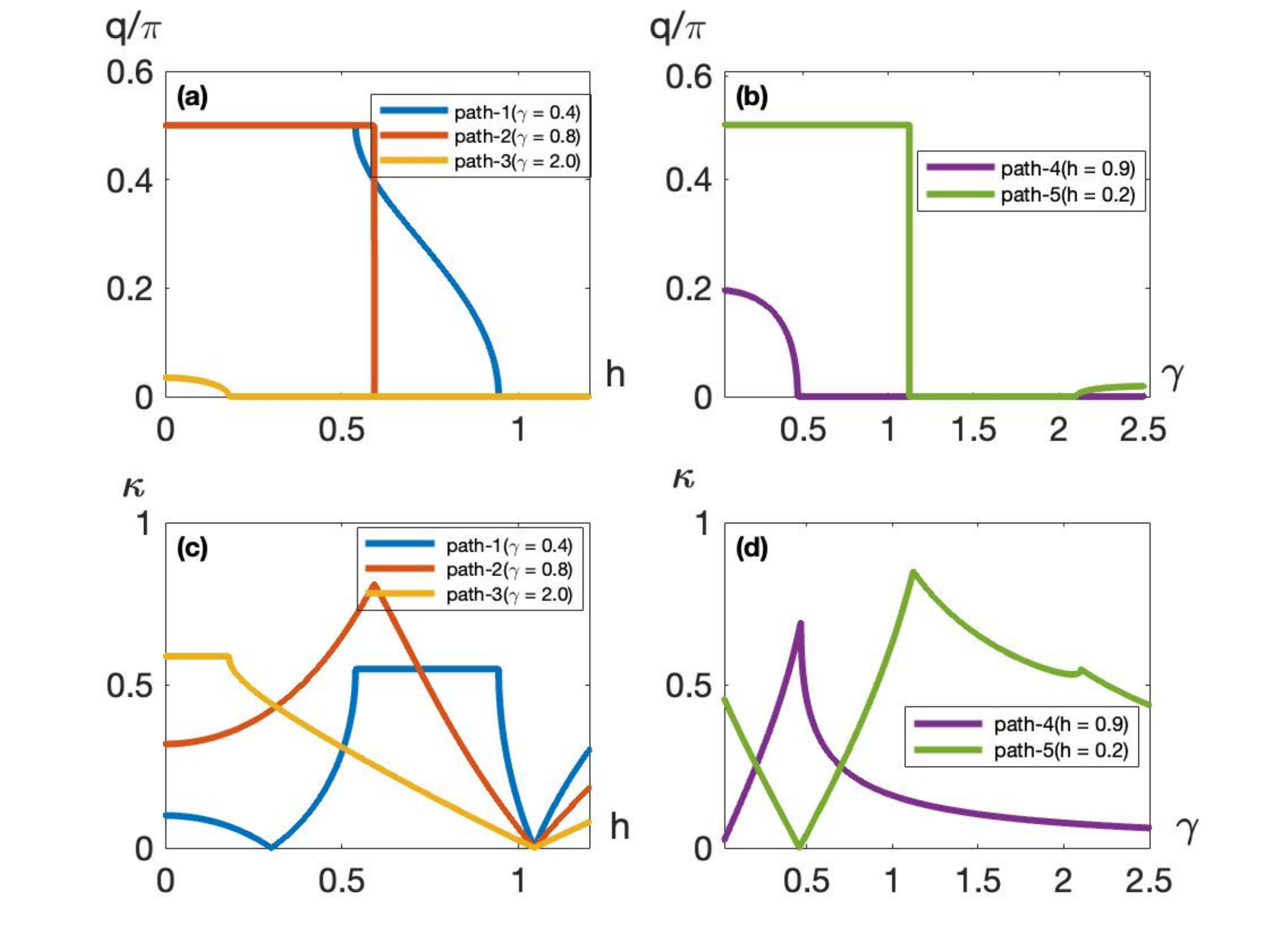}}
\caption{The modulation wave number ($q$) and the inverse correlation length ($\kappa$)  along several paths indicated in ($h,\gamma$) plane,  see Fig.~\ref{HaDelGam} for the general case $\delta \neq 0$ and $h_a \neq 0$. The disorder lines are accompanied by the cusps of $\kappa$, while the quantum phase transitions PM-FM and FM-$\mathcal{O}_z$ correspond to $\kappa=0$. The wave number is continuous, albeit not smooth on the disorder lines of the first kind, while it undergoes a discontinuity on the lines of the second kind.}
\label{KapQC}
\end{figure}

The qualitative mechanism for appearance of the DL2s for $h_a<\delta$ and $h_a>\delta$ is the same as explained in the end of the previous subsection after Eq.~\eqref{qDel}. The DL2 correspond to a jump of the wave vector of oscillations $0 \leftrightarrow \pi/2$ and a cusp of the correlation length, see Figs.~\ref{ZpmC} and \ref{KapQC}. The regions of the IC oscillations are shown in green on the phase diagram in Fig.~\ref{HaDelGam}.
The wave vector $q$ and the inverse correlation length $\kappa$ calculated along several paths shown in Fig.~\ref{HaDelGam}
are plotted  in Fig.~\ref{KapQC}. The qualitative behavior of the correlation functions is similar to that plotted in Fig.~\ref{CFA} for different regions of the phase diagram.

\textit{To summarize the new results of this section:} addition of the dimerization and the staggered field brings about quite rich structure of the DLs
on the phase diagram of the model, richer than the simple DL circle $h^2+\gamma^2=1$ \cite{McCoyII:1971} for the known case $h_a=\delta=0$.
The DLs are controlled by the behavior of zeros of the partition function, which become, in the limit $T \to 0$, the zeros $(z_\pm)$ of the spectrum.
Two kinds of DLs are found in the model:

(I). The DL of the first kind (or simply DL) corresponding to the merging complex conjugate roots $z_+=z_-$. In this case the wave number of the IC oscillations evolves smoothly in the oscillating region where $z_+=z_-^\ast$.

(II). The DL of the second kind (DL2) occurs at the point $z_+=-z_-$, where $z_\pm \in \mathbb{R}$. The asymptotes of the correlation functions are controlled
by the root $z_\sharp$ closest to the unit circle:
\begin{equation}
\label{zcl}
  \min \{||z_+|-1|, ||z_-| -1| \}  \longmapsto  z_\sharp~.
\end{equation}
DL2 separates the regions without oscillations ($q=0$) and with the commensurate ($q=\pi/2$) oscillations.  The discontinuity of the wave number
at the DL2 can be written as:
\begin{equation}
\label{dq}
  q = \frac{\pi}{4} (1-\mathrm{sign}   z_\sharp )~.
\end{equation}
A cusp in the correlation length occurs for the both kinds of DLs.

%
%
%
%
%
\section{Ground state factorization}\label{Factor}
%
%
%
%
%
The systematic method for the search of disentangled (factorized) states in quantum chains was proposed by  M\"{u}ller and
coworkers \cite{Muller:1982,*Muller:1985}.
The properties of separable states  in various spin models have been actively studied in the literature,
see, e.g. \cite{Amico:2006,Giorgi:2009,Rossignoli:2009,*Cerezo:2017,*Rossignoli:2020,Illuminati:2008,*Illuminati:2009,Jones:2021}
and more references in there. The problem has many facets and potential applications, and one of the outstanding
physical questions is the nature of the disentanglement points: can it be viewed as a quantum transition of
some kind? See, e.g, \cite{Wolf:2006,Amico:2006,Illuminati:2013,WeiGold:2005,*WeiGold:2011}

Very recently \cite{Chitov:2021} the factorization of the GS of the $XY$ chain in the uniform transverse field
was related to a special class of the complex conjugate zeros $z_+=z_-^*$ of its partition function (in the limit $T\to0$).
In that model the parametric line of factorization coincides with the disorder line $\gamma^2 +h^2 =1$ \cite{McCoyII:1971,Franchini:2017},
where the roots $z_\pm$ merge, since their imaginary parts vanish.
The disorder line (factorization) transition is a very weak feature, and it is not straightforward to classify it in the standard Ehrenfest scheme. For the quantum   $XY$ chain it was rigorously shown \cite{Maciazek:2016} that its GS energy is smooth and even infinitely differentiable function on the disorder (factorization) line. The gap does not close at those points, and probably the only clean-cut nonanalytic feature is its cusp.
\footnote{\label{PTRuelle}Some time ago Ruelle \cite{Ruelle:1974} proposed an alternative definition of phase transition without invoking non-analyticities of thermodynamic potentials. It is defined as a discontinuity in the qualitative properties of correlation functions. Discontinuity in the derivative of the correlation length in case of DLs qualifies then as a transition in Ruelle's classification.
Moreover, for the DL as a modulation transition with the modulation wave vector continuously growing deep into
the IC phase, one can use the notion of the critical index of modulation $\nu_{\s L}$, defined as $q \propto |h_{\s DL} - h|^{\nu_{\s L}}$
(with $ \nu_{\s L}=1/2$) by Nussinov and co-workers \cite{Nussinov:2012}.}

The goal of this section is to reveal relation between the DLs of the model \eqref{XYHam} analyzed in Sec.~\ref{DL3} and the factorized states
yet to be detected. To deal with the latter we follow the method of M\"{u}ller and coworkers \cite{Muller:1982,*Muller:1985} with some modifications.
Their original idea was to rotate each spin of the chain in the $xz$ plane to make the transformed Hamiltonian ferromagnetic with the fully
separable (factorized) GS. In the Hamiltonian \eqref{XYHam} with $\delta=0$ we rotate spins by the position-dependent angle $\vartheta_n$:
\begin{equation}
\label{Rot}
\left(\begin{array}{c}
\tilde{S}_{n}^{x} \\
\tilde{S}_{n}^{z}
\end{array}\right)=U_{n}^{-1}\left(\begin{array}{c}
S_{n}^{x} \\
S_{n}^{z}
\end{array}\right)
U_{n}=
\left(\begin{array}{cc}
\cos 2 \vartheta_{n} & \sin 2 \vartheta_{n} \\
-\sin 2 \vartheta_{n} & \cos 2 \vartheta_{n}
\end{array}\right)\left(\begin{array}{l}
S_{n}^{x} \\
S_{n}^{z}
\end{array}\right)
\end{equation}
with the rotation matrix
\begin{equation}
\label{Urot}
  U_{n}=\left(\begin{array}{cc}
\cos \vartheta_{n} & \sin \vartheta_{n} \\
-\sin \vartheta_{n} & \cos \vartheta_{n}
\end{array}\right)
\end{equation}
The transformed Hamiltonian
\begin{equation}
\label{Htil}
  \tilde{H}=U^{-1} H U =H(\tilde{S}),~~U \equiv \prod_{n=1}^{N} U_n
\end{equation}
can be brought to a desired form with the choice of two angles of rotation on the even and odd sites:
\begin{equation}
\label{Thetan}
 \vartheta_n=
\left\{
\begin{array}{lr}
\vartheta_e~, ~~n=2l \\[0.2cm]
\vartheta_0~, ~~n=2l-1\\
\end{array}
\right.
\end{equation}
Assuming an even number of sites $N$ and periodic boundary conditions $S_{N+1}^{\alpha}=S_{1}^{\alpha}$,
the transformed Hamiltonian written in terms of the original spins reads
\begin{widetext}
\begin{equation}
\label{HamRot}
   \tilde{H} =- \sum_{n=1}^{N}~ \big[ (1-\gamma) \big( S_{n}^{x }S_{n+1}^{x} + S_{n}^{y} S_{n+1}^{y} \big)
   + \sin 2\vartheta_e \sin 2\vartheta_o  S_{n}^{z} S_{n+1}^{z}  \big]
   +\sum_{l=1}^{N/2}~ \big[ (h-h_a) \cos  2\vartheta_o S_{2l-1}^{z}
   +  (h+h_a) \cos  2\vartheta_e S_{2l}^{z} \big]~,
\end{equation}
\end{widetext}
where the other terms of $\tilde{H}$ (not written explicitly) vanish in its GS if the conditions
\begin{equation}
\label{CosCos}
  \cos  2\vartheta_e \cos  2\vartheta_o =\frac{1-\gamma}{1+\gamma}
\end{equation}
and
\begin{eqnarray}
\label{CondMix1}
  (1+\gamma) \sin 2 \vartheta_e \cos 2 \vartheta_o &=&(h-h_a) \sin 2 \vartheta_o \\
\label{CondMix2}
  (1+\gamma) \sin 2 \vartheta_o \cos 2 \vartheta_e &=&(h+h_a) \sin 2 \vartheta_e
\end{eqnarray}
are satisfied. The solution of Eqs.~(\ref{CosCos},\ref{CondMix1},\ref{CondMix2}) exists for
$h^2+ \gamma^2=1+h_a^2$, thus the factorization occurs on the DL circle (see Fig.~\ref{HaGam}).
Introducing the factorization field $h_f$, we find
\begin{equation}
\label{hf}
  h_f=h_{\s DL}^{(1)}=\pm \sqrt{1+h_a^2-\gamma^2}~,
\end{equation}
while the other DL field $h_{\s DL}^{(2)}=\pm h_a$ does not solve Eqs.~(\ref{CosCos},\ref{CondMix1},\ref{CondMix2}),
thus $h_f \neq h_{\s DL}^{(2)}$.
The factorizing field \eqref{hf} and the angles of spin rotation are related as
\begin{equation}
\label{CosEO}
  \cos^2 2 \vartheta_{e/o}= \frac{(\sqrt{1+h_a^2-\gamma^2} \pm h_a)^2+(1-\gamma)^2}{
  (\sqrt{1+h_a^2-\gamma^2} \pm h_a)^2+(1+\gamma)^2 }~,
\end{equation}
These angles in their turn are related to the roots \eqref{zpmHa} merging on the DL circle as
\begin{equation}
\label{ThetaZpm}
  z_{+} =z_{-} \equiv z_{\s DL} = \cos  2\vartheta_e \cos  2\vartheta_o~.
\end{equation}

\begin{figure*}[t]
\centering{\includegraphics[width=13.0cm]{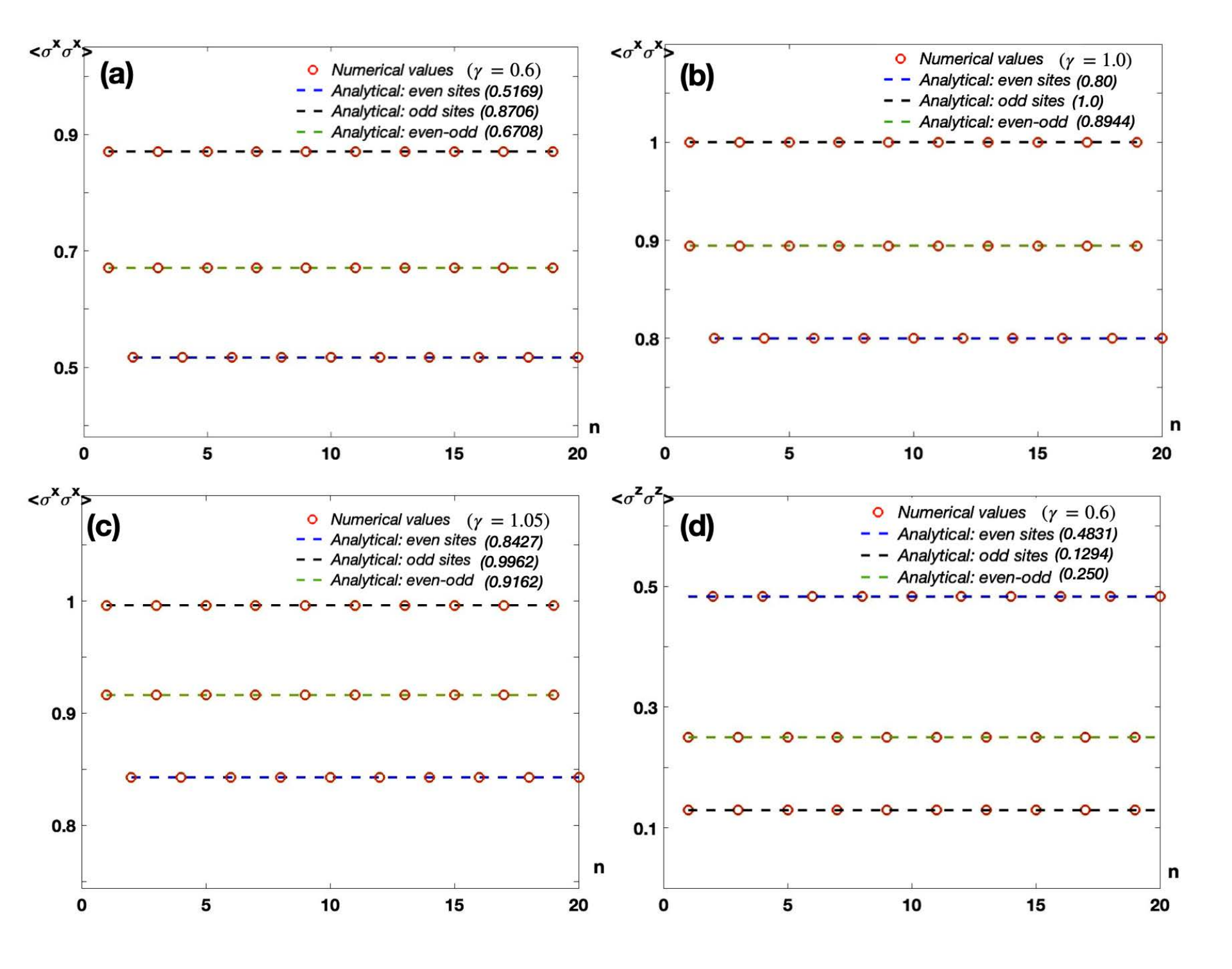}}
\caption{ Constant spin-spin correlation functions at three points on the circle of disentanglement \eqref{hf}. Panels (a) and (d): $\langle \sigma _m^x \sigma_m^x \rangle$ and $\langle \sigma _m^z \sigma_m^z \rangle$ at a point on the boundary between regions (B) and (C) of the FM phase, see Fig.~\ref{HaGam}; (b): point of intersection of two DLs: \eqref{hf} and $h=h_a$; (c): point on the boundary between (D) and IC oscillations.}
\label{DisentCorr}
\end{figure*}

\begin{figure}[]
\centering{\includegraphics[width=8.0cm]{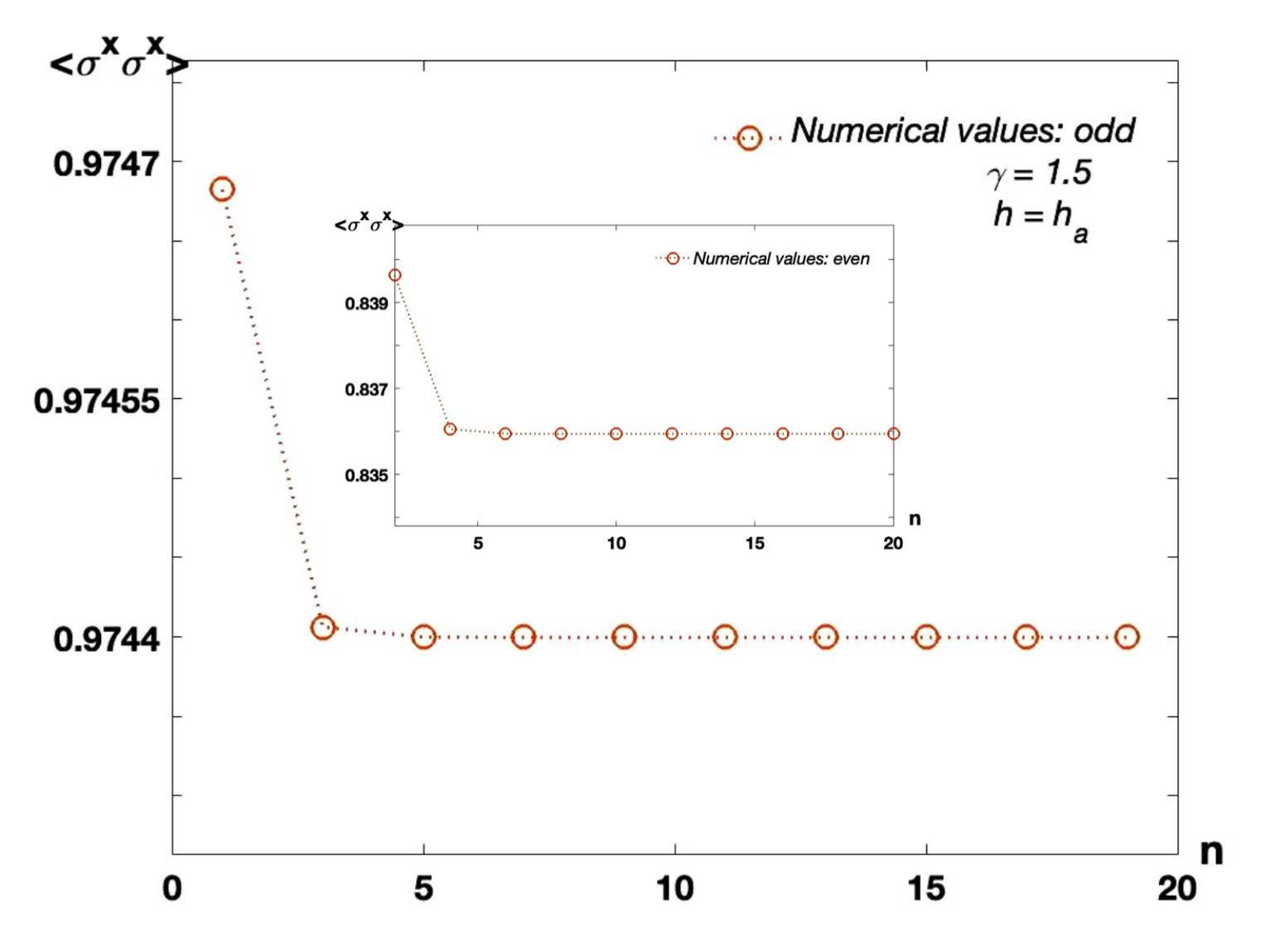}}
\caption{ Spin-spin correlation function $\langle \sigma _m^x \sigma_m^x \rangle$ on the disorder line $h=h_a$ where the GS is not disentangled.
Contrary to the case shown in Fig.~\ref{DisentCorr}, the correlation functions are not constant.}
\label{haCorr}
\end{figure}

The Hamiltonian \eqref{HamRot} is the $XXZ$ chain with the interaction parameter ($\Delta \equiv J_z/J$)
\begin{equation}
\label{DeltaR}
  \Delta =  \frac{\sin 2\vartheta_e \sin 2\vartheta_o}{1-\gamma}~.
\end{equation}
generated by the transformation of the original non-interacting model \eqref{XYHam}.
This $XXZ$ chain is subjected to the uniform and staggered transverse fields $\tilde{h}$ and $\tilde{h}_a$, respectively,
defined as
\begin{eqnarray}
\label{RenH1}
   \frac{(h-h_a) \cos  2\vartheta_o}{1-\gamma} &\equiv& \tilde{h} -\tilde{h}_a \\
\label{RehH2}
   \frac{(h+h_a) \cos  2\vartheta_e}{1-\gamma} &\equiv& \tilde{h} +\tilde{h}_a~.
\end{eqnarray}
The above equations together with Eq.~(\ref{CosCos}) lead to:
\begin{equation}
\label{Hcr}
  \tilde{h}^2=1 +\tilde{h}_a^2.
\end{equation}
An important conclusion then follows from the above equations and the rigorous results
\cite{Alcaraz:1995} for the $XXZ$ chain: with the fields satisfying \eqref{Hcr} and $\Delta>0$ \eqref{DeltaR}, the GS of $\tilde{H}$
\eqref{HamRot} is ferromagnetic with the state ket:
\begin{equation}
\label{FM}
|\mathrm{FM} \rangle=\prod_{n=1}^{N}|\uparrow\rangle_{n}=\prod_{n=1}^{N}\left(\begin{array}{l}1 \\ 0\end{array}\right)_{n}~.
\end{equation}
From \eqref{Htil} the GS of $H$ is obtained as
\begin{equation}
\label{Psi}
 |\Psi \rangle = U |\mathrm{FM} \rangle ~.
\end{equation}
Note that the expression on the right hand side of Eq.~\eqref{CosEO} is positive and less than one, so $\cos 2 \vartheta_\sharp$ is real, while its
sign depends on whether $\gamma<1$ or $\gamma>1$, cf. Eq.~\eqref{CosCos}. Assuming $\gamma<1$, we can localize the solutions in the range
$ |2\vartheta_\sharp| \leq \pi/2$. So, for $\cos 2 \vartheta_\sharp$ we choose the positive branch of the square root in Eq.~\eqref{CosEO}, as well as for
$ \cos \vartheta_\sharp =\sqrt{(1+\cos 2 \vartheta_\sharp)/2}$, while
we allow two signs of $ \sin \vartheta_\sharp =\pm \sqrt{1-\cos^2 \vartheta_\sharp}$ to account for the two-fold degeneracy of the GS.
(Here $\sharp =e/o$ for even/odd sites, respectively.)
Then
\begin{widetext}
\begin{equation}
\label{Psipm}
  {\left| \Psi _{\pm}  \right\rangle} =\prod _{l=1}^{N/2}
  \big( \cos \vartheta_o \left| \uparrow  \right\rangle_{2l-1}  \pm \sin \vartheta_o \left| \downarrow \right\rangle_{2l-1}   \big)
  \big( \cos \vartheta_e \left| \uparrow  \right\rangle_{2l}  \pm \sin \vartheta_e \left| \downarrow \right\rangle_{2l}   \big),
\end{equation}
\end{widetext}
with
\begin{equation}
\label{GSProd}
  \left\langle \Psi _{\pm } \Psi _{\pm } \right\rangle =1,~~
  \left\langle \Psi _{+} \Psi _{-} \right\rangle =\Big( \frac{1-\gamma}{1+\gamma} \Big)^{N/2}~.
\end{equation}
The factorized state \eqref{Psipm} is maximally disentangled. To put it more quantitatively let us invoke the two-site concurrence
$\mathcal{C}$ introduced by Wootters \cite{Wootters:1998} as a measure of entanglement. It can be calculated as \cite{Chitov:2021}
\begin{equation}
\label{C}
 \mathcal{C}=\sum _{m\ne n}{\left\langle \Psi  \right|} i\sigma _{m}^{y} i\sigma _{n}^{y} {\left| \Psi  \right\rangle} ~.
\end{equation}
For the states \eqref{Psipm} one can easily verify:
\begin{equation}
\label{YY}
 {\left\langle \Psi _{\pm}  \right|} \sigma _{m}^{y} {\left| \Psi _{\pm}  \right\rangle} =
  {\left\langle \Psi _{\pm}  \right|} \sigma _{m}^{y} \sigma _{n}^{y} {\left| \Psi _{\pm}  \right\rangle} =0,~~\forall~ m \neq n.
\end{equation}
thus  $\mathcal{C}=0$ on the factorized DL circle \eqref{hf}, as expected \cite{Amico:2006}.
To obtain the longitudinal magnetization we find
\begin{equation}
\label{Mx}
 {\left\langle \Psi _{\pm}  \right|} \sigma _{m}^{x} {\left| \Psi _{\pm}  \right\rangle} =
 \pm \sin 2 \vartheta_\sharp~.
\end{equation}
Thus two degenerate ground states $\left| \Psi _{\pm}  \right\rangle$
can be linked to two possible orientations of the spontaneous order parameter $m_x$. Similarly,
\begin{equation}
\label{Mz}
 {\left\langle \Psi _{\pm}  \right|} \sigma _{m}^{z} {\left| \Psi _{\pm}  \right\rangle} =
  \cos 2 \vartheta_\sharp~.
\end{equation}
A hallmark of the ground state factorization are the constant correlation functions.
For the disentangled state they are found in a closed form:
\begin{equation}
\label{XXZZ}
\left\langle \sigma _{m}^{\alpha} \sigma _{n}^{\alpha} \right\rangle =
\left\langle \sigma _{m}^{\alpha} \right\rangle \left\langle \sigma _{n}^{\alpha} \right\rangle~,
~~\forall~ m \neq n~,
\end{equation}
with the average values $\langle \sigma _{m}^{\alpha} \rangle$ given by Eqs.~(\ref{Mx},\ref{Mz},\ref{CosEO}).
The above analytical results are in agreement with the direct numerical calculations of the correlation functions via the Toeplitz
determinants with the explicit formulas provided in \cite{Chitov:2019}. The representative plots of those functions for three points on the line of disentanglement are shown in Fig.~\ref{DisentCorr}. For comparison we also show in Fig.~\ref{haCorr} the spin-spin correlation function on the disorder line $h=h_a$ where the GS is not disentangled.


The behavior of correlators on the generic DLs and on the DLs with the factorized disentangled GSs is  stemming from the analytical
properties of the generating function \cite{Chitov:2021,Jones:2021,Jones:2021B}. The elements of the Toeplitz determinants yielding
various spin-spin and string-string correlation functions are the two-point Majorana correlators:
\begin{equation}
\label{Gab}
  \langle i a_n b_m \rangle = \oint _{\left|z\right|=1}\frac{dz}{2\pi i}  z^{n-m-1} \frac{\mathcal{R}(z)}{E_+(z) E_-(z)}~,
\end{equation}
where the explicit formula for $\mathcal{R}(z)$ can be found in \cite{Chitov:2019}. On the DL two complex conjugate zeros of the spectrum $z_\pm$
merge into a single real degenerate root $z_{\s DL}$, rendering its contribution to the integrand of \eqref{Gab} meromorphic:
\begin{equation}
\label{GabDL}
  \langle i a_n b_m \rangle \overset{\mathrm{\s DL}}{=} \oint _{\left|z\right|=1}\frac{dz}{2\pi i}  z^{n-m} \frac{\tilde{\mathcal{R}}(z)}{(z-z_{\s DL})(z-z_{\s DL}^{-1})}~,
\end{equation}
For the limiting case $h_a=\delta=0$ the aforementioned restructuring of singularities of the generating function in \eqref{GabDL} in sufficient to both:

\textit{a)} yield the DL as a boundary of the modulated region with IC oscillations, and 
 
\textit{b)} render the appropriate Toeplitz determinants size-independent on the whole DL, such that Eqs.~\eqref{XXZZ} hold and the state is disentangled (factorized) \cite{Chitov:2021}.

\textit{An important conclusion from the results of this section} is that occurrence of DL as a point where the complex roots $z_\pm$ merge does not necessarily lead to a factorized ground state, i.e., disentanglement.  It is true for the case $h_a=\delta=0$, but turning on the staggered field $h_a \neq 0$ already changes the situation.  As we have shown, for the factorizing field $h_f$ \eqref{hf} and the DL field $h_{\s DL}^{(2)}=\pm h_a$ the properties \textit{a)} and \textit{b)} do not coincide, however the DL circle \eqref{hf} is disentangled.
One can check that in the presence of dimerization $\delta \neq 0$, there are no solutions of the equations for $\tilde{H}$ with the factorized GS.
Thus, $\forall~ \delta \neq 0$, the states on the DLs analysed in Sec.~\ref{DL3} for the cases B ($h_a=0$) and C ($h_a \neq 0$) are entangled, in agreement with general arguments of \cite{Illuminati:2008,*Illuminati:2009,Giorgi:2009}.

%
%
%
%
%
\section{Zero-energy Majorana edge states}\label{MES}
%
%
%
%
%

Let us first find the 1D winding number (or the Pontryagin index) defined as \cite{WenZee:1989,SchnyderRyu:2011}
\begin{equation}
\label{Nw}
  N_w=\frac{1}{2\pi i} \int_{BZ} dk \mathrm{Tr}[\partial_k \ln \hat{D}]~,
\end{equation}
where $\hat{D}$ is introduced in Eq.~\eqref{Dk}. It can be readily written as
\begin{equation}
\label{NwCompl}
    N_w=\oint _{\left|z\right|=1} \frac{dz}{2\pi i} \partial_z \ln  \mathrm{det}\hat{D}~.
\end{equation}
Then the topological number \eqref{NwCompl} is the logarithmic residue of $\mathrm{det}\hat{D}$.
It accounts for the excess of the number of zeros over the number of poles (weighted with their degrees of multiplicity)
of $\mathrm{det}\hat{D}$ inside the unit circle on the complex plane \cite{Whittaker:1915}.
The zeros of $\mathrm{det}\hat{D}$ are also zeros of the spectra, cf. Eqs.~({\ref{EpmDD},\ref{detD}).
Any change of winding number occurs only when a root (roots) crosses the unit circle $|z|=1$, which can happen only at
a point of quantum phase transition \cite{Verresen:2018,Chitov:2021}. By the same token, $N_w$ does not change upon
crossing disorder lines.

From  Eq.~(\ref{detD}) and known root positions with respect to the complex unit circle
(see Figs.~\ref{ZpmA},\ref{ZpmB},\ref{ZpmC}), we readily obtain the winding numbers for the three gapped phases of the model \eqref{XYHam}:
\footnote{\label{NwOld} The phase of logarithm is not uniquely defined. In earlier related work \cite{Chitov:2019,Chitov:2020} the limit of the phase of logarithm in Eq.~\eqref{Nw} at the ends of the BZ was  chosen  such that $N_w$ differed from the current result \eqref{NwAll} by 1. Note that \eqref{Nw} is
defined mod 2. Definition \eqref{NwCompl} is robust and more intuitively transparent.}
\begin{equation}
\label{NwAll}
N_w=
  \left\{
    \begin{array}{l}
      \phantom{\pm} 0,~\mathrm{PM~phase} \\[0.2cm]
      \mp 1, ~\mathrm{FM~phase},~ \gamma \gtrless 0\\[0.2cm]
      \phantom{\pm} 0, ~\mathcal{O}_z(\pi/2)~ \mathrm{phase.}  \\
    \end{array}
   \right.
\end{equation}

In the search of the zero-energy Majorana edge states of the model \eqref{XYHam} we will follow the earlier related papers
\cite{Karevski:2000,DeGottardi:2011,DeGottardi:2013b,Chitov:2018}. We introduce two zero-energy Majorana operators $\hat \alpha_0$
and  $\hat \beta_0$ as
\begin{equation}
\label{AB0}
2 \hat \eta_{\s 0}^\dag \equiv \hat \alpha_{\s 0} +i \hat \beta_{\s 0}  =\sum_{n=1}^N \big[\phi_n \hat a_n + i \psi_n \hat b_n \big],~\{ \phi_n,\psi_n \} \in \mathbb{R},
\end{equation}
where $\hat \eta_{\s 0}^\dag$ denotes the creation operator of the Bogoliubov fermion with zero energy. (The Hamiltonian is diagonal in terms of
the Bogoliubov fermions). The involution of Majorana operators amounts to the requirement of the wave function normalization in \eqref{AB0}:
\begin{eqnarray}
  \label{ANorm}
  \hat \alpha_{\s 0}^2 &=& \sum_{n=1}^N \phi_n^2=1~, \\
  \hat \beta_{\s 0}^2 &=& \sum_{n=1}^N \psi_n^2=1~.
 \label{BNorm}
\end{eqnarray}
The Heisenberg equations for $\hat \alpha_{\s 0}$ and $\hat \beta_{\s 0}$ yield
\begin{equation}
\label{Commut}
  [\hat \alpha_{\s 0}, \hat H]=[\hat \beta_{\s 0}, \hat H]=0~.
\end{equation}
Writing the Hamiltonian \eqref{XYFermi} with open boundary conditions ($N=2L+1$) via the Majorana operators \eqref{Maj}, one can bring the commutativity condition \eqref{Commut} to the form of the iterative equation for the Majorana wave function
\begin{equation}
\label{phiT}
  \left(\begin{array}{l}
   \phi_{n+1} \\
   \phi_n
  \end{array}\right)
=\hat T_n
  \left(\begin{array}{l}
   \phi_n \\
   \phi_{n-1}
  \end{array}\right)
\end{equation}
with the transfer matrix
\begin{equation}
\label{Tn}
\hat T_n =\left(
          \begin{array}{cc}
            \frac{2h_n}{t_n+ \gamma } & -\frac{t_{n-1}-\gamma}{t_n+\gamma} \\[0.2cm]
            1 & 0 \\
          \end{array}
        \right)
\end{equation}
where
\begin{equation}
\label{hntn}
  h_n \equiv h +(-1)^n h_a,~~t_n \equiv 1 +(-1)^n \delta~.
\end{equation}
For the other wave function we get
\begin{equation}
\label{psiT}
  \left(\begin{array}{l}
   \psi_{n+1} \\
   \psi_n
  \end{array}\right)
=\hat T_n(-\gamma)
  \left(\begin{array}{l}
   \psi_n \\
   \psi_{n-1}
  \end{array}\right)~.
\end{equation}
There are only two distinct matrices $\hat T_n$:
\begin{equation}
\label{Tpm}
  \hat T_{\s 2l/2l-1}=\hat T_{\pm}.
\end{equation}
Using the period 2 transfer matrix
\begin{equation}
\label{T2}
  \hat T \equiv \hat T_+ \hat T_- =
  \left(
          \begin{array}{cc}
            \frac{4(h^2-h_a^2)-(1-\delta)^2+\gamma^2}{(1+\gamma)^2-\delta^2}
            &  -\frac{2(h+h_a)(1+\delta-\gamma)}{(1+\gamma)^2-\delta^2} \\[0.2cm]
              \frac{2(h-h_a)}{1-\delta+\gamma} & -\frac{1+\delta-\gamma}{1-\delta+\gamma} \\
          \end{array}
        \right)
\end{equation}
the solution of \eqref{phiT} can be written as
\begin{equation}
\label{phiN}
  \left(\begin{array}{l}
   \phi_{\s 2m+1} \\
   \phi_{\s 2m}
  \end{array}\right)
=\hat T^m
  \left(\begin{array}{c}
   \phi_{\s 1} \\
    0
  \end{array}\right)~.
\end{equation}
Similarly,
\begin{equation}
\label{psiN}
  \left(\begin{array}{l}
   \psi_{\s 2m+1} \\
   \psi_{\s 2m }
  \end{array}\right)
=\hat T^m(-\gamma)
  \left(\begin{array}{c}
   \psi_{\s 1} \\
    0
  \end{array}\right)~.
\end{equation}
To look for the tentative Majoranas localized near the right end of the chain, we write equations \eqref{Commut}
in the different iterative form as
\begin{equation}
\label{phiTR}
  \left(\begin{array}{l}
   \phi_{n-1} \\
   \phi_n
  \end{array}\right)
=\hat T_n(-\gamma, -\delta)
  \left(\begin{array}{l}
   \phi_n \\
   \phi_{n+1}
  \end{array}\right)~,
\end{equation}
leading to the solution
\begin{equation}
\label{phiNR}
  \left(\begin{array}{l}
   \phi_{\s 2L-2m+1} \\
   \phi_{\s 2L-2m+2}
  \end{array}\right)
=\hat T^m(-\gamma)
  \left(\begin{array}{c}
   \phi_{\s 2L+1} \\
    0
  \end{array}\right)~.
\end{equation}
In the same way we obtain
\begin{equation}
\label{psiNR}
  \left(\begin{array}{l}
   \psi_{\s 2L-2m+1} \\
   \psi_{\s 2L-2m+2}
  \end{array}\right)
=\hat T^m
  \left(\begin{array}{c}
   \psi_{\s 2L+1} \\
    0
  \end{array}\right)~.
\end{equation}

To advance further \cite{Chitov:2018} we write the transfer matrix $\hat T$
via two orthogonal idempotent operators (projectors)
\begin{equation}
\label{Ppm}
  \hat{\mathcal{P}}_\pm \equiv \pm \frac{ \hat T -\lambda_\mp \mathbb{1} }{\lambda_+ - \lambda_-}
\end{equation}
as
\begin{equation}
\label{TLpm}
  \hat T =\lambda_+ \hat{\mathcal{P}}_+  + \lambda_- \hat{\mathcal{P}}_- ~,
\end{equation}
where $\lambda_\pm$ are two eigenvalues of $\hat T$.
Then
\begin{equation}
\label{Tpower}
  \hat T^m =\lambda_+^m  \hat{\mathcal{P}}_+  + \lambda_-^m \hat{\mathcal{P}}_-
\end{equation}
As follows from Eqs.~(\ref{ANorm},\ref{BNorm},\ref{phiN},\ref{psiN},\ref{phiNR},\ref{psiNR},\ref{Tpower}), for the existence of the normalizable
zero-energy Majorana states in the thermodynamic limit $N \to \infty$, the eigenvalues of the transfer matrix must lie inside the complex unit circle
$|\lambda_\pm|<1$.

The key result is that the eigenvalues of the transfer matrix $\hat T$ \eqref{T2} are zeros of the spectrum \eqref{zpm}:
\begin{equation}
\label{lamz}
  \lambda_\pm= z_\pm~,
\end{equation}
in agreement with the earlier findings \cite{Chitov:2018} for the case $h_a=\delta=0$. Also, according to Eq.~\eqref{zinv}, the eigenvalues
of $\hat T(-\gamma)$ are $z_\pm^{-1}$.
\textit{Thus the zeros of the spectrum $z_\pm$, which are also the zero-temperature limit of the partition function zeros, control the critical points, modulation transitions, disentanglement, topological numbers, and the zero-energy Majorana states.} The properties on the \textit{edge} ($\hat T$) are controlled by the \textit{bulk} parameters ($z_\pm$).

Note that counting of the localized Majorana modes from number of zeros of the resolvent of the transfer matrix $\hat T$ within the unit circle on the complex plane  \cite{DeGottardi:2011,DeGottardi:2013b} is still valid, however it does not provide an independent information after the eigenvalues of the transfer matrix are being identified as zeros of the spectrum \eqref{lamz}.

Since the information on the roots $z_\pm$ \eqref{zpm} is available for all regions of the model's phase diagram
(see Figs.~\ref{ZpmA},\ref{ZpmB},\ref{ZpmC}), we can give the qualitative answers about existence of the Majorana states prior of any calculations:

1). PM and $\mathcal{O}_z(\pi/2)$ phases:  Among two roots $z_\pm$  one is always located inside the unit circle, and the other one outside (the same for their reciprocals $z_\pm^{-1}$), thus none of the four equations (\ref{phiN},\ref{psiN},\ref{phiNR},\ref{psiNR}) has a normalizable solution, and the zero-energy Majorana modes do not exist in those phases.

2). FM phase with $m_x \neq 0$ ($\gamma >0$): The solution of (\ref{phiN}) for the wave function $\phi_n$ of the Majorana zero-energy state $\hat \alpha_{\s 0}$ localized near the left end of the chain and the solution of (\ref{psiNR}) for $\psi_n$
of the Majorana state $\hat \beta_{\s 0}$ localized near the right end, exist. The zero-energy Bogoliubov (Dirac) fermion $\hat \eta_{\s 0}^\dag$ \eqref{AB0} is a
superposition of two Majoranas localized on the opposite ends of the chain.

3). FM phase with $m_y \neq 0$ ($\gamma <0$): $\phi_n \leftrightarrow \psi_n$ since in this case equations (\ref{psiN}) and (\ref{phiNR}) have normalizable solutions.  The zero-energy Majoranas $\hat \alpha_{\s 0}$  and $\hat \beta_{\s 0}$ are now localized near the right and left ends of the chain, respectively.

From the above equations we readily find the explicit formulas for the wave function of the $\hat \alpha_{\s 0}$ Majorana in the $m_x$-FM phase:
\begin{eqnarray}
\label{phiOdd}
  \phi_{ 2m+1} &=& \left( \hat T^m \right)_{ 11} \phi_{ 1}~, \\
   \left( \hat T^m \right)_{ 11} &=& \frac{(z_+^m-z_-^m) T_{ 11}-(z_+^{m-1}-z_-^{m-1})z_+ z_-  }{z_+ - z_-}~,  \nonumber \\
  \phi_{ 2m} &=& \left( \hat T^m \right)_{ 21} \phi_{ 1}~, \\
  \label{phiEven}
    \left( \hat T^m \right)_{21} &=& \frac{(z_+^m-z_-^m) T_{ 21}}{z_+ - z_-} ~,\nonumber
\end{eqnarray}
where the matrix elements $T_{11}$ and $T_{21}$ are given in \eqref{T2}.
The normalization condition \eqref{ANorm} yields
\begin{equation}
\label{phi1}
  \phi_1 = \Big[1+\sum_{m=1}^L \Big\{ \left( \hat T^m \right)_{ 11}^2+ \left( \hat T^m \right)_{ 21}^2 \Big\} \Big]^{-1/2}
\end{equation}
For the wave function of the $\hat \beta_{\s 0}$ Majorana we get
\begin{eqnarray}
\label{psiEven}
  \psi_{ 2L-2m+1} &=& \left( \hat T^m \right)_{ 11} \psi_{ 2L+1}~, \\
  \psi_{2L- 2m+2} &=& \left( \hat T^m \right)_{ 21} \psi_{ 2L+1}~,
  \label{psiOdd}
\end{eqnarray}
with $\psi_{ 2L+1}$ determined by the right hand side of Eq.~(\ref{phi1}).

The wave functions are exponentially decaying with the penetration depth (inverse correlation length) determined by the root $z_\pm$ closest to the unit circle. In the light yellow regions without oscillations
(see Figs.~\ref{ZpmA},\ref{ZpmB},\ref{ZpmC}) the roots are positive, while the negative root closest to the unit circle results in the four-periodic oscillations of the wave functions in the light blue regions of the phase diagram.
In the IC oscillating (light green) regions of the phase diagram where $z_+=z_-^*$ (cf. notations \eqref{kC}) the formulas become particularly simple:
\begin{eqnarray}
\label{phiOddIC}
 \left( \hat T^m \right)_{ 11} &=& e^{-2 \kappa (m-1)} \frac{\sin 2qm}{\sin2q} T_{11} \nonumber \\
 &-&e^{-2 \kappa m} \frac{\sin 2q(m-1)}{\sin2q} \\
  \left( \hat T^m \right)_{ 21} &=& e^{-2 \kappa (m-1)} \frac{\sin 2qm}{\sin2q} T_{21}~.
  \label{phiEvenIC}
\end{eqnarray}
The plots of the wave functions for representative cases are shown in Fig.~\ref{MajoranaWF}

\begin{figure}[]
\centering{\includegraphics[width=7.0cm]{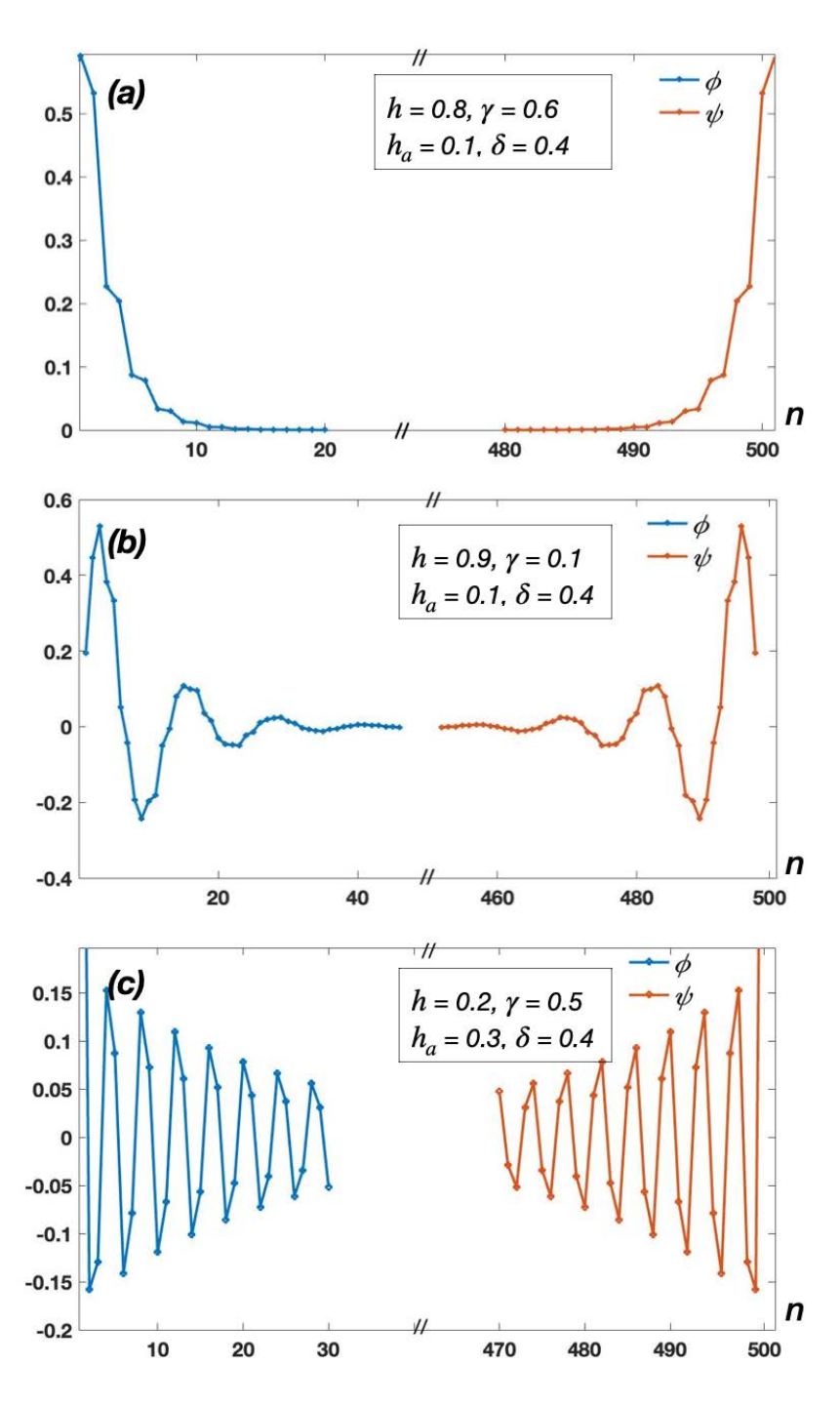}}
\caption{Wave functions $\phi, \psi$ of the zero-energy  Majoranas $a$, $b$ localized near the left and right ends of the chain, respectively.
The functions are  calculated in three regions (B), (C), and (D) of the $m_x$-FM phase (see Fig.~\ref{HaDelGam}) with $N=501$.
Panel (a): region (B) without oscillations; (b) region of the IC oscillations (C) with the wave number $q =0.1583 \pi$; (c): region (D) with four-periodic
oscillations $q=\pi/2$.}
\label{MajoranaWF}
\end{figure}
%
%

%
%
%
%
%
\section{Conclusion}\label{Concl}
%
%
%

In this work we present a detailed study of the exactly solvable dimerized ferromagnetic $XY$ chain with uniform and staggered transverse fields,
which is equivalent in the fermionic representation to the noninteracting dimerized Kitaev-Majorana chain with modulated chemical potential.
The model has three known gapped phases with local and nonlocal (string) orders, along with the gapless incommensurate (IC) phase
in the $U(1)$ limit. It is shown that the model along with  the continuous quantum phase transitions possesses weaker singularities known
as disorder lines (DLs) or modulation transitions. The latter, reported for the first time in this model, are localized in the ferromagnetically
ordered regions of the phases diagram.  The DLs are shown to occur in this model in two types: DLs of the first kind with continuous appearance
of the IC oscillations,  and DLs of the second kind (DL2) corresponding to a jump of the wave number which controls the oscillations.
 A cusp in the correlation length occurs for the both kinds of DLs.

It is also shown that this model hosts the zero-energy Majorana edge modes in its ferromagnetic phases.
In each phase there is a couple of the Majorana states localized on the opposite ends of the chain, which can be viewed as a ``deconfined"
Bogoliubov (Dirac) fermion with zero energy. The exactly calculated wave functions of those edge Majoranas are shown to acquire modulations on the DLs.

It is demonstrated that the DLs are the locus of the separable ground states. More specifically, the ground state is proved to be separable (factorized)
on a subset of the DLs (circle) in the case of zero dimerization $\delta=0$. In this state the model is disentangled, which is remarkably manifested by
constant correlation functions. With the explicit formula for the factorized ground state, the correlation functions are calculated exactly analytically, in
agreement with the numerical calculations of the Toeplitz determinants.

From the view point of the general theory of phase transitions, probably the most important advancement made in the present study is that all the above results are obtained from analysis of the properties of zeros of model's partition function, analytically continued onto the complex wave numbers. In the ground state they evolve into complex zeros $z_\pm$ of the spectrum of the Hamiltonian, such that: 

1). Those zeros signal a quantum critical transition when one of the roots $z_\pm$ (or both simultaneously) cross the complex unit circle, $|z_\pm |=1$. In this case (only!) the topological winding number changes as well.

2). Merging of the complex conjugate roots $z_+=z_-$
signals the DL (continuous modulation transition) which can also be accompanied by the ground state factorization (disentanglement). 
The DL of the second kind (DL2) with discontinuous modulation and without disentanglement, occurs when $z_+=-z_-$ and $z_\pm \in \mathbb{R}$.

3). The roots $z_\pm$ are also shown to be the eigenvalues of the transfer matrix \eqref{lamz}, thus their number inside the unit circle on the complex plane controls also the existence of the localized zero-energy edge Majoranas.

We think a very important direction of future analysis is to get a deeper insight on the analytical properties of generic disorder lines to single out their subsets where the ground state is factorized, the model is disentangled, and the appropriate Toeplitz determinants are size-independent.

~\\

\noindent $^\dagger$ We deeply regret the sudden death of our co-author, Pavel Nikolaevich Timonin (23/11/1948 - 20/03/2022),
during the final stage of the work on this paper. We did our best for the manuscript to accurately represent his vision of the
results and to meet his very high standards.

%
\begin{acknowledgments}
G.Y.C. thanks N.G. Jones for helpful correspondence and especially C. Bourbonnais for his support.
G.Y.C. gratefully acknowledges financial support from Institut quantique (IQ) of
Universit\'{e} de Sherbrooke and Regroupement qu\'{e}b\'{e}cois sur les mat\'{e}riaux de pointe (RQMP).
K.G. thanks Girish Sharma and Salvatore R. Manmana for useful discussions and gratefully  acknowledges funding by
the Deutsche Forschungsgemeinschaft (DFG, German Research Foundation) - 217133147/CRC 1073, project B03 and computational
resources provided by the IIT Mandi.
P.N.T. was supported by the Ministry of Education and
Science of the Russian Federation through the state assignment in the field of scientific
activity, project No.~0852-2020-0032 (BAS0110/20-3-08IF).
\end{acknowledgments}
%


\bibliography{C:/Papers/BibRef/CondMattRefs}
%
%
%
\end{document}